\documentclass[10pt,journal,compsoc]{IEEEtran}
\usepackage[T1]{fontenc}
\ifCLASSOPTIONcompsoc

\usepackage{cite}

\usepackage{bm}
\usepackage{amsmath}
\usepackage[table]{xcolor}
\definecolor{verylightgray}{rgb}{0.85, 0.85, 0.85}

\usepackage{amsthm}
\usepackage{float}
\usepackage{setspace}
\usepackage{amssymb}
\usepackage{stfloats}
\usepackage{cite}
\usepackage{ragged2e}
\usepackage{colortbl}
\usepackage[ruled,vlined,linesnumbered]{algorithm2e}
\usepackage{algorithmic}
\usepackage{amsfonts}
\usepackage{mathrsfs}
\usepackage{amsmath,amsthm}
\usepackage{array,booktabs}
\usepackage{subfigure}
\usepackage{multirow}
\usepackage{cuted}
\usepackage{multicol}
\usepackage{graphicx}
\usepackage{placeins} 
\usepackage{subfigure}
\usepackage{graphicx,xcolor,bm}
\usepackage{hyperref}
\usepackage{threeparttable}
\usepackage{dcolumn}
\usepackage{setspace}
\usepackage{makecell}
\usepackage{lipsum}
\usepackage{enumerate}
\usepackage{hhline}

\usepackage{tabularx}

\newtheorem{remark}{Remark}

\newtheorem{definition}{Definition}

\usepackage{cite,bm}
\graphicspath{{figures/}}
\usepackage{bm}

\usepackage[protrusion=true,expansion=true]{microtype}
\usepackage{caption}
\usepackage[font=footnotesize,labelfont=bf]{caption}

\makeatletter

\begin{document}
	
	\title{Auctioning Future Services in Edge Networks with Moving Vehicles: $N$-Step Look-Ahead Contracts for Sustainable Resource Provision}
	
	\author{Ziqi Ling, Minghui Liwang, \IEEEmembership{Senior Member}, \IEEEmembership{IEEE}, Xianbin Wang, \IEEEmembership{Fellow}, \IEEEmembership{IEEE}, \\
		Seyyedali Hosseinalipour, \IEEEmembership{Senior Member}, \IEEEmembership{IEEE}, Zhipeng Cheng, \IEEEmembership{Member}, \IEEEmembership{IEEE},\\ Sai Zou, \IEEEmembership{Senior Member}, \IEEEmembership{IEEE}, Wei Ni, \IEEEmembership{Fellow}, \IEEEmembership{IEEE}, Xiaoyu Xia, \IEEEmembership{Senior Member}, \IEEEmembership{IEEE}
		\thanks{Ziqi Ling (lingziqi@stu.xmu.edu.cn) is with the School of Informatics, Xiamen University, Fujian, China. Minghui Liwang (minghuiliwang@tongji.edu.cn) is with the Department of Control Science and Engineering, Shanghai Institute of Intelligent Science and Technology, the National Key Laboratory of Autonomous Intelligent Unmanned Systems, and also with Frontiers Science Center for Intelligent Autonomous Systems, Ministry of Education, Tongji University, Shanghai, China. Xianbin Wang (xianbin.wang@uwo.ca) is with the Department of Electrical and Computer Engineering, Western University, Ontario, Canada. Seyyedali Hosseinalipour (alipour@buffalo.edu) is with Department of Electrical Engineering, University
at Buffalo–SUNY, NY, USA. Zhipeng Cheng
(chengzp\_x@163.com) is with the School of Future Science and Engineering, Soochow University, Jiangsu, China. Sai Zou (dr-zousai@foxmail.com) is with the College of Big Data and Information Engineering, Guizhou University, Guiyang, China. Wei Ni (Wei.Ni@ieee.org) is with Data61, CSIRO, Australia.
 Xiaoyu Xia (xiaoyu.xia@rmit.edu.au) is with the School of Computing Technologies, RMIT University, Melbourne, Australia.}}
	
	

	\IEEEtitleabstractindextext{
	\begin{abstract}
		\justifying
		Timely resource allocation in Edge-assisted networks with moving vehicles is critical for supporting compute-intensive vehicular services such as autonomous driving and real-time navigation. In these environments, vehicles generate resource demands that can quickly exceed the capacity of individual edge servers (ESs). To address this, ESs frequently borrow and lend resources among each other using market-based strategies, with double auctions serving as one of the most prominent mechanisms for enabling such resource exchanges. However, effectively coordinating this resource trading in spatio-temporal dynamic vehicular settings presents significant challenges  due to two intertwined limitations:
		\textit{(i)} the spatio-temporal unpredictability of resource needs arising from vehicle mobility, and
		\textit{(ii)} the latency overhead of real-time double auctions.
		To overcome these limitations, we propose a look-ahead contract-based auction framework that shifts/migrates decision-making from \textit{runtime} to \textit{planning time}. The key idea behind our method is to establish, in advance, $N$-step future service contracts between ESs using demand forecasts and modified double auction mechanisms. Specifically, we design a two-stage system: in the former stage,  \textit{(i)} a long short-term memory (LSTM)-based prediction module estimates future resource needs and determines ES roles (buyer or seller) across $N$ time frames, and \textit{(ii)} a pre-double auction mechanism generates multi-slot contracts that specify traded resource quantities, prices, and penalties. Then, in the latter stage, these  contracts are enforced in real-time without re-running auctions.
		Our framework explicitly integrates energy consumption, transmission costs, and contract breach risks into utility models, enabling formation of truthful, individually rational, and energy-aware contracts. Experiments on real-world (UTD19) and synthetic traces show that our method achieves notable improvement in time efficiency, energy use, and social welfare compared to existing benchmark methods.
	\end{abstract}
	
	}
	\maketitle
	\IEEEdisplaynontitleabstractindextext
	%
	\IEEEpeerreviewmaketitle
	
	\section{Introduction}
	\IEEEPARstart{R}{ecent} advances in artificial intelligence (AI) and their integration into mobile networks have accelerated the development of intelligent vehicles, enabling services such as autonomous driving, traffic monitoring, and real-time navigation \cite{ref6}. However, these AI-driven services are highly resource-intensive, while on-board vehicular computing remains constrained by limited processing power and energy \cite{ref1}, creating a mismatch between demand and capability. A common solution is computation offloading to remote cloud or nearby edge servers (ESs) \cite{ref7,ref8,ref9,ref10}, giving rise to edge computing where resources are deployed closer to vehicles \cite{ref2}. Yet, the spatio-temporal dynamics of vehicular workloads often cause resource imbalance across ESs, e.g., overloaded servers in dense traffic regions versus underutilized ones in sparse areas. This highlights the need for dynamic and cooperative resource sharing among ESs to effectively handle fluctuating vehicular demands \cite{ref3}.

	As a natural consequence, \textit{incentivizing inherently selfish ESs to engage in collaborative resource sharing becomes essential. This necessitates the establishment of suitable economic frameworks, including efficient payment mechanisms and service-level agreements, to facilitate financially viable service trading among ESs.} Among various approaches, \textit{the double auction}\footnote{A double auction is a market mechanism in which multiple buyers and sellers concurrently submit their bids and asks, respectively. A clearing price is then determined to facilitate efficient matching between supply and demand. 
	 } \cite{ref12, Profit Maximization Incentive Mechanism for Resource Providers in Mobile Edge Computing, auction5,auction1} has emerged as a prominent cornerstone for orchestrating resource exchanges between deficit and surplus ESs. In this market-driven model, ESs lacking sufficient capacity function as buyers, while those with idle resources act as sellers. By dynamically matching supply and demand of ESs through competitive bidding, the double auction not only delivers economically advantageous outcomes for all ESs but also enhances the overall attractiveness of the edge computing ecosystem to the moving vehicles \cite{A Truthful Auction for Graph Job Allocation in Vehicular Cloud-Assisted Networks}.
	\subsection{Primary Motivation}
	By embedding double auction mechanisms into edge networks with moving vehicles, geographically distributed ESs can flexibly exchange resources on a pay‐per‐use basis. Nevertheless, there are two interrelated challenges that can hinder its practical deployment, which are of our particular interest:
	
	\noindent
	\textit{(Challenge 1) Spatio-temporal demand variability: }
	Vehicular traffic and computing workloads fluctuate sharply across locations and time, causing ESs to alternate between resource deficit and resource surplus states. This volatility complicates \textit{(i)} role selection (buyer vs. seller) in each auction round and \textit{(ii)} accurate multi-frame forecasting of resource quantities to procure or offload. For example, an ES may need to purchase extra compute at 09:00 to handle a morning surge but possess excess compute capacity to sell by 13:00.
	
	
	\noindent
	\textit{(Challenge 2) Decision-making overhead of complex interactions among ESs: }Double auctions require aggregating many concurrent bids and asks to determine a clearing price and settle payments. As the market scales, iterative bidding, participant matching, and price computation become coordination-heavy and latency-prone. In highly dynamic IoV scenarios, these delays can break service continuity, e.g., vehicle requests may time out or vehicles may leave an ES's coverage before auction results are available  \cite{Bridge the Present and Future: A Cross-Layer Matching Game in Dynamic Cloud-Aided Mobile Edge Networks}.

	\textit{Addressing the aforementioned two challenges serves as our primary motivation in this work.} To this end, we design a novel trading framework where each ES can seamlessly switch between buyer and seller roles across successive auction rounds (time intervals). To achieve low latency and energy efficiency, the framework integrates three key components:
	
	
	\noindent \textit{(i) Adaptive Resource Prediction:} A lightweight prediction mechanism that estimates the dynamic demands of each ES across time frames, enabling timely role selection (buyer or seller) in upcoming auctions.
	
	\noindent \textit{(ii) Pre-Double Auction with Multi-Period Contracts:} A contract-based auction scheme that secures agreements for the next $N$ time frames, specifying resource quantities, payments, and breach penalties. Energy-awareness is explicitly embedded to incentivize sustainable and efficient resource sharing.
	
	\noindent \textit{(iii) Contract Enforcement:} A dedicated process that synchronizes actual per frame exchanges with both pre-signed contracts and real-time vehicular demands, ensuring reliable service delivery, energy and cost efficiency, and fair compensation among ESs.
%
%
	
	From a timeline perspective, the first two components are executed prior to actual resource exchanges and jointly constitute the first stage of our framework. Also, the third one operates in real-time to accommodate instantaneous service demands, forming the second stage of the framework.

\vspace{-2mm}
	\subsection{Relevant Literature Investigations}
	This section reviews related literature while highlighting this work's key distinctions and contributions.


    \vspace{-2mm}
	\subsubsection{Resource Sharing among Edge Servers} 
	Addressing the resource scarcity of ESs has become a central challenge in edge computing, motivating extensive research on resource provisioning and task scheduling  \cite{Reliability-aware Optimization of Task Offloading for UAV-assisted Edge Computing}. Much of the existing work in this domain emphasizes cloud-edge collaboration \cite{ref11, ref4} and cloud-edge-user frameworks \cite{ref5, ceu, ceu2}, which orchestrate multi-tier resources to optimize latency and energy consumption. While these studies focus on vertical resource management, there is an increasing interest in horizontal resource sharing among ESs. For example, \textit{Liu} et al. \cite{share3} proposed a multi-stage resource sharing mechanism across edge service providers to improve resource utilization, and \textit{Zafari} et al. \cite{share4} modeled ES cooperation via games to maximize individual profits. Beyond profit, recent works target social welfare: \textit{Yue} et al. \cite{ref3} introduced alliance formation among ESs, while \textit{Vera-Rivera} et al. \cite{share2} leveraged blockchain to address security and privacy in cross-ES resource allocation and scheduling.
	

	While the above studies have advanced collaborative edge computing, they often neglect spatio-temporal fluctuations in resource demand caused by vehicle mobility and time-varying workloads. \textit{To fill this gap, we propose a prediction-driven, mutually beneficial resource-trading paradigm that anticipates each ES's future demand and enables economically fair exchanges of compute and storage, thereby addressing the operational realities of mobile-edge ecosystems.}

    \vspace{-1mm}
	\subsubsection{Auction-based Resource Sharing/Trading}
	Auctions have been widely adopted to enable resource sharing and trading \cite{auction1, auction2, auction3, auction5, auction7}. For instance, \textit{Kang} et al. \cite{auction5} proposed an iterative combinatorial double auction to support task offloading in cloud-edge collaborative networks, while \textit{Huang} et al. \cite{auction7} introduced a neutral third-party auctioneer to coordinate interactions and avoid direct bargaining between service and infrastructure providers. Crucially, auctions are designed to satisfy a set of desired economic properties \cite{auction3}, fostering active and trustworthy participation. For example, \textit{Ren} et al. \cite{auction6} formulated an auction-driven resource market that maximizes system utility while guaranteeing truthfulness (i.e., participants are incentivized to bid true valuations) and individual rationality (i.e.,participants gain non-negative utility).
	
	
	While the aforementioned auction-based schemes have advanced the state-of-the-art in edge resource allocation, they largely neglect the substantial overhead incurred by the bidding process of auctions, which can undermine responsiveness in highly dynamic network scenarios. \textit{To mitigate this, we investigate a novel auction-based resource sharing framework with a forward-looking, contract-driven pre-auction stage: ESs negotiate and formalize resource-sharing agreements ahead of real-time demands. By pre-establishing these contracts, the system circumvents much of the traditional auction latency, yielding faster resource allocation.}

    \vspace{-1mm}
	\subsubsection{Pricing Rules in Resource Sharing/Trading Markets}
	Effective pricing schemes (i.e., determining buyers' payments and sellers' earnings) are crucial for economic efficiency in resource-sharing environments. Static pricing approaches \cite{price2, price3, price4} explore equilibrium models, privacy preservation, and fairness incentives but struggle with rapidly changing demand. This has motivated dynamic pricing frameworks \cite{dynamic2, dynamic3, dynamic4, price1}  that adjust prices in real-time, improving flexibility and system welfare. For example, \textit{Baek} et al. \cite{dynamic2} analyzed fairness and efficiency of dynamic edge pricing via game theory, \textit{Tang} et al. \cite{dynamic3} incorporated overbooking to boost utilization and profits, and subsequent studies \cite{dynamic4,price1} used game-theoretic strategies to enhance economic efficiency and system performance.

	Although the above studies have proposed novel pricing schemes, real-time negotiation and optimization associated with these schemes can introduce latency, hindering latency-sensitive applications in dynamic edge environments. \textit{To overcome this, we integrate a dynamic pricing mechanism within our pre-auction framework, where contractual terms and prices are set in advance. This eliminates on-site negotiation, enabling low-latency resource exchanges and uninterrupted service delivery.}
 \begin{figure}[!t]
		\centering
		\includegraphics[width=\linewidth]{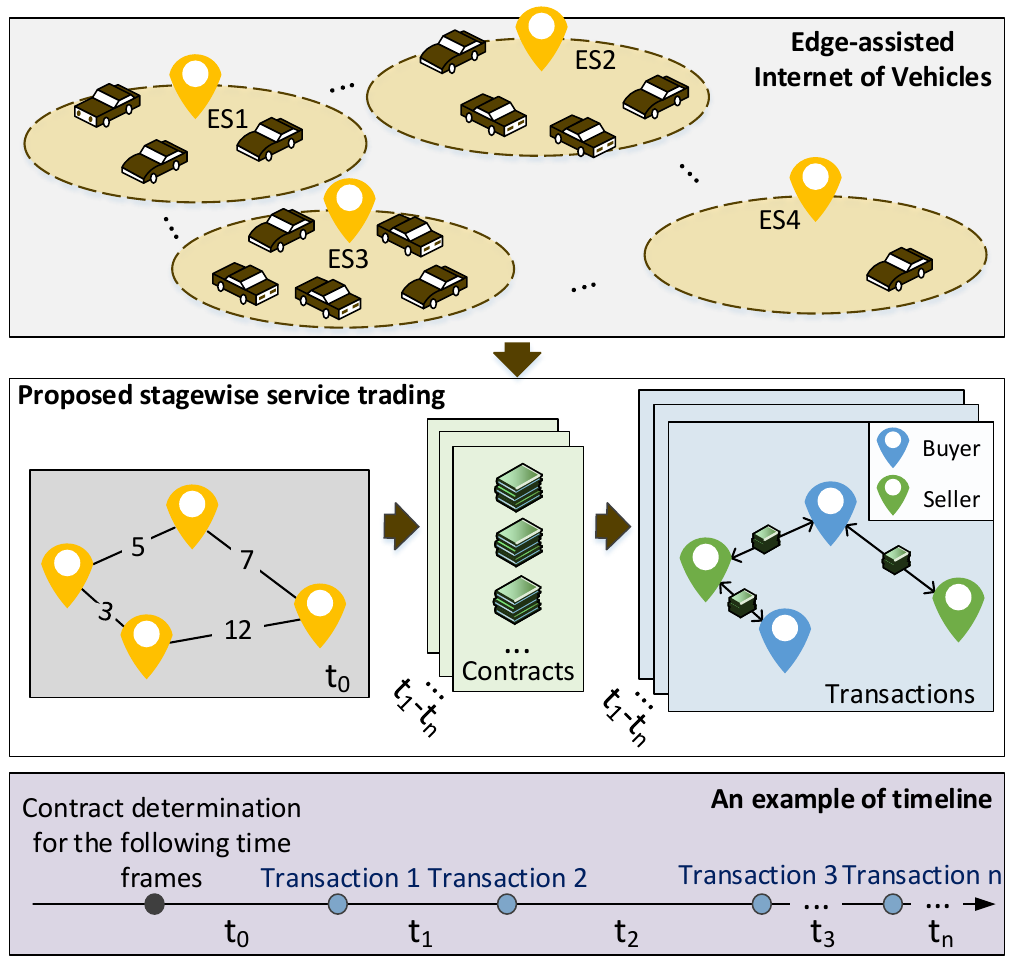}
        \vspace{-6.2mm}
		\caption{Framework of our consideration (the upper box), the diagram of service trading via $N$-step ahead for green contracts (the middle box), where the numbers 3, 5, 7, and 12 indicate the abstract distances between ESs, and an example timeline (the bottom box).
		}
		\label{fig:fig1}
	\vspace{-0.45cm}
	\end{figure}   

    \vspace{-5mm}
	\subsection{Novelty and Contribution}
	To our knowledge, \textit{this paper is the first attempt among the existing literature  to design a stagewise resource sharing paradigm across geographically distributed ESs that integrates (i) resource usage prediction, (ii) $N$-step look-ahead contract-based planning of future transactions, and (iii) guaranteed contract execution based on real-time network conditions.} By partitioning the resource sharing decisions across discrete time frames and allowing each ES to fluidly assume buyer or seller roles, our framework enables many-to-many\footnote{In this paper, a many-to-many resource sharing model is adopted, where each ES acting as a buyer can procure resources from multiple sellers, while an ES acting as a seller can simultaneously provide resources to multiple buyers.} resource sharing that dynamically aligns the ESs' resource supply with the resource demand generated by moving vehicles. Our key contributions can be summarized as follows:
	
	\noindent $\bullet$ This study proposes a novel framework for dynamically trading time-varying computing resources among ESs in an Edge-assisted Internet of Vehicles (EdgeIoV) environment, where vehicle mobility and fluctuating workloads create unpredictable demand. Unlike prior work assuming static roles or relying on real-time negotiations, our approach enables each ES to flexibly switch between buyer and seller roles across successive time frames, guided by intelligent forecasts of future resource demand and supply. By explicitly modeling role uncertainty and transactional complexity, our framework preemptively coordinates resource exchanges to mitigate congestion and underutilization.
	
	\noindent
	$\bullet$ To ensure responsiveness and energy efficiency, we formulate a long-term optimization problem that maximizes the expected social welfare of all ESs over the planning horizon. The problem incorporates cumulative utilities, pre-established contractual commitments, and penalties for contract breaches. Given its NP-hard nature, arising from combinatorial prediction of future demand and multi-period contract design, we develop a dedicated prediction mechanism for dynamic vehicular edge workloads and integrate it into the pre-auction trading process.
	
	\noindent
	$\bullet$ We propose a stagewise framework for efficient and reliable resource trading in dynamic EdgeIoV environments. First, an intelligent prediction model estimates each ES’s future resource usage and assigns its role (buyer or seller) for upcoming time frames. Based on these predictions, a double-auction mechanism forms pre-signed contracts between prospective buyers and sellers, eliminating latency from real-time bidding. To handle IoV uncertainties, a contract revocation mechanism allows buyers to withdraw from part of the agreements when conditions change, enhancing flexibility and robustness in real-world deployments.
	
	\noindent
	$\bullet$ We provide a formal analysis demonstrating that the proposed stagewise design guarantees key economic properties, including truthfulness and individual rationality, thereby fostering trust and stability within the market.
	
	\noindent $\bullet$ We evaluate our approach using both real-world vehicular traffic datasets and synthetic numerical datasets to assess its effectiveness in realistic EdgeIoV scenarios. Across key metrics (including time efficiency, energy efficiency, and social welfare) and varying network scales, our method consistently outperforms existing benchmarks.
 \vspace{-0.4cm}
\section{Overview and Modeling Framework}
\setlength{\abovedisplayskip}{4pt}
\setlength{\belowdisplayskip}{4pt}

\subsection{Overview }
We consider a service trading market in an EdgeIoV environment, empowered by an auction scheme that performs $N$-step look-ahead planning for future resource trading. Specifically, ESs are collected by the set $\mathcal{ES} = \{ ES_1, \dots,ES_m,\dots, ES_{|\mathcal{ES}|} \}$ and can provide computing services to vehicles within their coverage areas. To improve resource coordination among geographically distributed ESs and attract more vehicles with better service quality, we define \textit{two roles} for the ESs, buyers and sellers, within a \textit{role-adaptive} framework that supports dynamic switching over time. In this market, buyers may engage with multiple sellers, and sellers may serve multiple buyers, forming a many-to-many (M2M) matching.

 To facilitate seamless/timely service provisioning in  a dynamic EdgeIoV environment, we decompose the resource trading process into \textbf{two distinct stages}. The \textbf{first stage} conducts a pre-double auction based on \textbf{predicted future resource demands}, encouraging ESs to sign \textbf{advance trading contracts} for upcoming time frames. The \textbf{second stage} focuses on the \textbf{execution of these pre-signed contracts}, enabling rapid service delivery during real-time operations.
 Specifically, the first stage occurs prior to practical resource transactions\footnote{A transaction represents a trading event, where sellers and buyers can share their resources and thus computing services by paying for certain expenses.}, where buyers and sellers report their estimated (i.e., predicted) resource supply and demand for the upcoming time frames, including their bids and asks, to the auctioneer. The auctioneer then coordinates a double auction based on this information and facilitates the generation of a series of look-ahead contracts
 (referred to as LAContracts for simplicity). Each of these contracts is implemented in practical transactions during the second stage, without the need for further negotiations. A schematic of our stagewise auction framework for EdgeIoV is illustrated in Fig. \ref{fig:fig1}.

\begin{figure}
	\centering
	\includegraphics[width=0.95\linewidth]{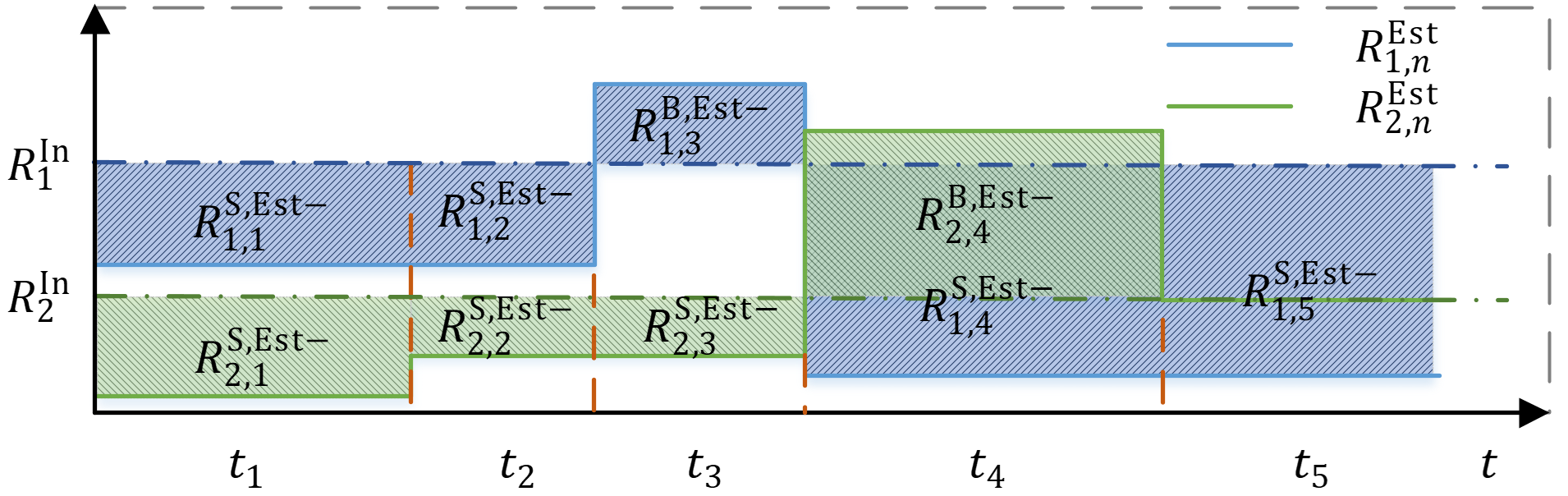}
    \vspace{-2mm}
	\caption{Example of role switching between two ESs over different time frames, where the label $\text{In}$ denotes the inherent resource, $\text{Est}$ denotes the estimated resource usage, and $\text{S,{Est-}}$ and $\text{B,{Est-}}$ denote the demand and the valuable resource, respectively (this symbol will be explained in following subsection).}
	
	\label{fig:fig2}
	\vspace{-0.5cm}
\end{figure}

Since the number of vehicles  under each ES varies over time (and thus resource demands as well as the roles of ESs, see Fig. \ref{fig:fig2}), we capture the system state over a set of time frames, gathered by the set $\bm{T} = \{ t_0, t_1, \dots, t_n, \dots, t_N \}$. Specifically, $t_0$ denotes the first stage, while practical transactions (during the second stage) start from $t_1$.  To capture the network dynamics \textit{the length of each time frame may vary} based on real-time factors such as traffic density, vehicle mobility, and network conditions. Importantly, since the resource demand covered by  ESs can change, an ES can act as a buyer in a transaction during $t_n$, while it can also serve as a seller in $t_{n+1}$ ($n>0$).

\vspace{-2mm}
 \subsection{Modeling of Buyers}
Recalling that an ES can have different roles across various transactions, we first denote the set of buyers during time frame \( t_n \) ($n>0$) as \( \mathcal{B}_n = \{ \bm{b}_{1,n}, \dots,\bm{b}_{i,n},\dots, \bm{b}_{|\mathcal{B}_n|,n} \} \), where \( \mathcal{B}_n \subseteq \mathcal{ES} \). In particular, each buyer \( \bm{b}_{i,n} \in \mathcal{B}_n \) is represented by a 6-tuple (with slight abuse of notation we consider the same notation for its corresponding 6-tuple) as  
$ \bm{b}_{i,n} = \{ \beta_{i,n}^\text{B}, \mathcal{R}_{i,n}^\text{B}, \mathsf{bid}_{i,n}', \omega_{i,n}^\text{B}, \eta_{i,n}^{\text{B},\text{Use}}, \eta_{i,n}^{\text{B},\text{Idle}} \}$, where \( \beta_{i,n}^\text{B} \) records the fixed identifier of this ES in set \( \mathcal{ES} \), e.g., \( \beta_{i,n}^\text{B} = 3 \) denotes $ES_3$ in set \( \mathcal{ES} \), helping distinguish different ESs with different roles.  \(\mathsf{bid}_{i,n}' \) describes the original unit price (e.g., per RB) a buyer is willing to pay, which is the same across all sellers, \( \omega_{i,n}^\text{B} \) indicates the unit cost of data transmission (over distance), which we consider as the cost of a buyer since they need to upload a certain amount of data when utilizing the seller’s resources for processing, \( \eta_{i,n}^\text{B} \) represents the unit cost of processing data locally, while \( \eta_{i,n}^{\text{B},\text{Idle}} \) represents the unit power consumption of a buyer when it has idle resources (e.g., stand-by power consumption). Also we use computing resource blocks (RB) (as also supported by \cite{RB}) to quantize the resources of ESs, while also facilitating the trading among them. In particular, we have  $
 \mathcal{R}_{i,n}^b = \{{R}_{i,n}^{\text{B},\text{Est}}, R_{i,n}^{\text{B},\text{Est-}}, R_{i,n}^{\text{B},\text{ln}}, R_{i,n}^{\text{B},\text{Act}}, R_{i,n}^{\text{B},\text{Tra}}\}
 $, 
 where ${R}_{i,n}^{\text{B},\text{Est}}$ describes the estimated demand of computing resources of buyer \( \bm{b}_{i,n} \) during time frame \( t_n \); \( R_{i,n}^{\text{B},\text{Est-}} = {R}_{i,n}^{\text{B},\text{Est}} - R_{i,n}^{\text{B},\text{ln}} \) describes the estimated resource shortage, where \( R_{i,n}^{\text{B},\text{ln}} \) represents the inherent computing resources owned by this ES. Note that when an ES becomes a buyer during \( t_n \), we usually have \( {R}_{i,n}^{\text{B},\text{Est}} > R_{i,n}^{\text{B},\text{ln}} \),  meaning that its resources may face difficulties to afford its demand. Besides, \( R_{i,n}^{\text{B},\text{Act}} \) denotes the actual resources required in the practical transaction during \( t_n \), and \( R_{i,n}^{\text{B},\text{Tra}} \) indicates the amount of resources specified in the LAContract during \( t_n \).
 
\vspace{-0.3cm}
\subsection{Modeling of Sellers}
Let \( \mathcal{S}_n = \{ \bm{s}_{1,n},  \dots,\bm{s}_{j,n},\dots, \bm{s}_{|\mathcal{S}_n|,n} \} \) represent the set of sellers during time frame $t_n$, where \(\mathcal{S}_n\subseteq \mathcal{ES} \), and \( \mathcal{B}_n \cap \mathcal{S}_n = \emptyset \). In particular, each \( \bm{s}_{j,n} \in \mathcal{S}_n \) (with slight abuse of notation) is represented by a corresponding 6-tuple: 
\(
\bm{s}_{j,n} = \{ \beta_{j,n}^\text{S}, \mathcal{R}_{j,n}^\text{S}, \mathsf{ask}_{j,n}, \eta_{j,n}^{\text{S},\text{Use}}, \eta_{j,n}^{\text{S},\text{Idle}}, q^\text{S}_{j,n}\}
\).
Here, \( \beta_{i,n}^\text{S} \) represents the inherent fixed identifier of an ES in set \( \mathcal{ES} \). Moreover, \( \mathsf{ask}_{j,n} \) represents the unit price of resources that a seller charges, and $q^\text{S}_{j,n}$ represents the per-unit penalty receivable by the seller when a breach occurs, while \( \eta_{j,n}^{\text{S},\text{Use}} \) and \( \eta_{j,n}^{\text{S},\text{Idle}}\) represent the unit power consumption incurred by  data analysis  and standby, respectively. Similar to buyers, \( R_{i,n}^\text{S} \) also represents a tuple as
\(
 \mathcal{R}_{j,n}^\text{S} = \{{R}_{j,n}^{\text{S},\text{Est}}, R_{j,n}^{\text{S},\text{Est-}}, R_{j,n}^{\text{S},\text{ln}}, R_{j,n}^{\text{S},\text{Act}}, R_{j,n}^{\text{S},\text{Tra}}\}
\),
where \({R}_{j,n}^{\text{S},\text{Est}} \) describes the estimated resource demand of seller \( \bm{s}_{j,n} \) during \( t_n \), \( R_{j,n}^{\text{S},\text{ln}} \) represents the inherent computing resources. When an ES becomes a seller during \( t_n \), we have 
\(R_{j,n}^{\text{S},\text{ln}} > {R}_{j,n}^{\text{S},\text{Est}} \), indicating that this seller may have surplus resources after covering its own demand, with  $R_{j,n}^{\text{S},\text{Est-}} = R_{j,n}^{\text{S},\text{ln}} - {R}_{j,n}^{\text{S},\text{Est}}$ denoting the number of available resources that can be purchased by ES buyers.
In addition, \( R_{j,n}^{\text{S},\text{Act}} \) denotes the actual resources required by \( \bm{s}_{j,n} \) during \( t_n \), and \( R_{j,n}^{\text{S},\text{Tra}} \) refers to the amount of resources specified in the LAContract for transaction during \( t_n \). 
 \vspace{-0.3cm}
\subsection{Modeling of Contracts}
During $t_0$, we aim to determine proper LAContracts among ESs for the following $N$ time frames, i.e., contracts for $t_1$ to $t_{|\bm{T}|}$.
To facilitate the auction for negotiating contracts for each $t_n$ $(n\geq1)$, we compactly represent the demand vector of all the buyers during $t_n$ as
\(
\bm{R}^\text{B}_n = [R^{\text{B},\text{Est-}}_{1,n}, R^{\text{B},\text{Est-}}_{2,n}, \dots,R^{\text{B},\text{Est-}}_{i,n},\dots, R^{\text{B},\text{Est-}}_{|\mathcal{B}_n|,n}].
\)
\textit{Note that in our designed auction, buyers will report their predicted resource requirements (i.e., $R^{\textnormal{B},\textnormal{Est-}}_{i,n}$), rather than the actual consumed resources (i.e., $R^{\textnormal{B},\textnormal{Act}}_{i,n}$).} This is because our auction is conducted prior to actual resource transactions (i.e., before practical demands occur), and thus it may not be practical for buyers to accurately specify their requirements during $t_n$. Consequently, we incorporate predicted demands to enable early bidding and buyer-seller matching, thereby ensuring efficient resource allocation (see Section \ref{sec:Methodology Design} for details).
Also, we compactly represent the original bidding vector of buyers for $t_n$, during $t_0$, as $\bm{B}'_n = [\mathsf{bid}'_{1,n}, \mathsf{bid}'_{2,n}, \dots, \mathsf{bid}'_{|\mathcal{B}_n|,n}]$. Note that different locations  of ESs can lead to various data transmission costs. Accordingly, the actual bid of each buyer $\bm{b}_{i,n}$ relies heavily on its distance to different sellers. Thus, we define the bid of $\bm{b}_{i,n}$ to seller $\bm{s}_{j,n}$ as
\vspace{-.5mm}
\begin{equation}\label{bid}
\mathsf{bid}_{i,j,n} = \mathsf{bid}'_{i,n} \left( 1 - \exp\Big(- \alpha / (\omega^\text{B}_{i,n} \, d_{i,j,n}) \Big) \right),
\end{equation}
Here, $\alpha > 0$ controls the rate of decay of the bid with respect to distance, $d_{i,j,n}$ denotes the distance between buyer $b_{i,n}$ and seller $\bm{s}_{j,n}$, and $\omega^\text{B}_{i,n} d_{i,j,n}$ represents overall transmission cost incurred by buyer $b_{i,n}$ when uploading data to seller $\bm{s}_{j,n}$. Subsequently, we construct the following bidding matrix across all buyer-seller pairs for each $t_n$:
	\(\bm{B}_n = \left[\mathsf{bid}_{i,j,n}\right]_{{i\in\{1,...,|\bm{\mathcal{B}_n}|\}},{j\in\{1,...,|\bm{\mathcal{S}_n}|\}}}.\)
In addition, we represent the sellers’ resource supply vector during time frame $t_n$  as 
\(
\bm{R}^\text{S}_n = [R^{\text{S},\text{Est-}}_{1,n}, R^{\text{S},\text{Est-}}_{2,n}, \dots, R^{\text{S},\text{Est-}}_{|\mathcal{S}_n|,n}]
\), 
and the vector of asking prices of sellers as 
\(
	\bm{A}_n = [\mathsf{ask}_{j,n}]_{j\in\{1,...,|\bm{\mathcal{S}_n}|\}}. 
\)
For each time frame $t_n\in \bm{T}$ $(n>0)$, we use \( \bm{\mathbb{C}}_n \) to denote the series of LAContract, which is expressed by 
\(
\bm{\mathbb{C}}_n = \left[\bm{C}_{i,j,n}\right]_{{i\in\{1,...,|\bm{\mathcal{B}_n}|\}},{j\in\{1,...,|\bm{\mathcal{S}_n}|\}}}
\)
where \( \bm{C}_{i,j,n} \) describes the contract between buyer \( b_{i,n} \) and seller \( \text{S}_{j,n} \) during \( t_n \), defined as \( \bm{C}_{i,j,n} = \{ \beta_{j,n}^\text{S}, \beta_{i,n}^\text{B}, R^\text{Con}_{i,j,n}, p^\text{B}_{i,n}, p^\text{S}_{j,n}, q_{i,n}^\text{B},q_{j,n}^\text{S}, c_{i,j,n} \} \). Here, \( R^\text{Con}_{i,j,n} \) represents the amount of trading resources stipulated in this contract, and \( p_{i,j,n} \) to denote the unit price (e.g., per RB) that the buyer pays. Also, we use \( q_{i,n}^\text{B} \) denotes the penalty that a buyer experiences for breaking the contract, e.g., when a buyer no longer needs to purchase its resources due to low demand\footnote{for the sake of analytical tractability and to better align with practical market behaviors, this paper adopts the assumption that only buyers bear the risk of contract default, while sellers are presumed to consistently fulfill their contractual obligations}. Thus, when buyers default on them, sellers are entitled to receive compensation, denoted as \( q_{j,n}^\text{S} \)\footnote{Within the market, sellers are constrained to trade only within the limits of their inherent resources, thereby guaranteeing fulfillment of contractual commitments and secure revenue acquisition. Conversely, buyers who procure resources in excess of their actual demand may exhibit opportunistic behavior by defaulting on their contracts.}. As ESs are geographically distributed, the varying distances between them result in different data transmission costs. 
We thus define the distance-induced cost per RB as \(c_{i,j,n}  =\mathsf{bid}'_{i,j,n} - \mathsf{bid}_{i,n} \), which represents the reduction in the payment received by seller $\bm{s}_{j,n}$ due to the transmission distance.
Subsequently, the total resources that may be purchased by buyer \( \bm{b}_{i,n} \), denoted by \( R_j^{\text{S},\text{Tra}} \), is the sum of resources traded with all the sellers, i.e., \( R^{\text{B},\text{Tra}}_{i,n} = \sum_{j=1}^{|\mathcal{S}_n|} R^\text{Con}_{i,j,n} \). The same setting  applies to sellers, where the total resources traded by seller \( \bm{s}_{j,n} \) during time frame \( t_n \) is given by: 
\(
R^{\text{S},\text{Tra}}_{j,n} = \sum_{i=1}^{|\mathcal{B}_n|} R^\text{Con}_{i,j,n}.
\)
If there is no contract established between buyer \( \bm{b}_{i,n} \) and seller \( \bm{s}_{j,n} \) during time frame \( t_n \), the values of \( R^\text{Con}_{i,j,n}\), \( p^\text{B}_{i,n}\), \( p^\text{S}_{j,n} \), \( q_{i,n}^\text{B} \), \( q_{j,n}^\text{S} \), \( c_{i,j,n} \) are all set to zero.

\vspace{-2mm}
\section{Look $N$ Steps Ahead for Sustainable Contracts over EdgeIoV}
In this section, we introduce a double auction mechanism for establishing suitable LAContracts to facilitate proactive service provisioning in the EdgeIoV environment. We first model the energy consumption associated with resource transactions, serving as the basis for evaluating the sustainability of the contracts.

\vspace{-2mm}
\subsection{Modeling of Energy Consumption}
To ensure the sustainability of a trading market, energy efficiency is crucial. Since energy consumption directly affects the utility of participating ESs, especially buyers,we first examine the energy cost incurred by buyers during resource transactions within a given time frame $t_n$. This cost constitutes a major part of their total expenditure and significantly influences the economic viability and sustainability of the proposed trading mechanism.
In particular, energy consumed by using the inherent resources of a buyer \( \bm{b}_{i,n} \) during \( t_n \) depends on the actual resource usage under its coverage, and is given by: 
\begin{equation}\label{energy1}
\hspace{-4mm}
	\resizebox{0.94\hsize}{!}{$
		\mathcal{E}^\text{B}_{i,n}
		{=} \Delta t_n \left(
		\underbrace{\min(R^{\text{B},\text{In}}_{i,n}, R^{\text{B},\text{Act}}_{i,n}) \eta^{\text{B},\text{Use}}_{i,n}}_{(a)}
		+
		\underbrace{\max \left( R^{\text{B},\text{In}}_{i,n} - R^{\text{B},\text{Act}}_{i,n}, 0 \right) \eta^{\text{B},\text{Idle}}_{i,n}}_{(b)}
		\right)
		$}\hspace{-0.4mm}.\hspace{-3.5mm}
\end{equation}

In \eqref{energy1}, \( \Delta t_n \) represents the duration of time frame \( t_n \); \textit{part (a)} describes the energy consumed by computing, while \textit{part (b)} denotes the standby energy consumption. Also, the energy consumed by a seller \(\bm{s}_{j,n}\), during \( t_n \), relies on the actual resource demand under its coverage and the resources used by the buyers it trades with, thus including two aspects:
\textit{(Aspect 1)} the energy consumed for data analysis, which is given by:
\begin{equation}
	E^{\text{S},1}_{j,n} = \Delta t_n \left( \min(R^{\text{S},\text{In}}_{j,n}, R^{\text{S},\text{Act}}_{j,n} + R^{\text{S},\text{Tra}}_{j,n} - \theta_{j,n}^\text{S}) \right)\eta^{\text{S},\text{Use}}_{j,n}, 	
\end{equation}
where \(  \theta_{j,n}^\text{S} \) represents the total number of resources defaulted by its contractual buyers during time frame \( t_n \). Note that the resources that can actually be utilized by the seller, represented by \( \min (R^{\text{S},\text{In}}_{j,n}, R^{\text{S},\text{Act}}_{j,n} + R^{\text{S},\text{Tra}}_{j,n} - \theta_{j,n}^\text{S})  \), cannot exceed its resource supply. And \textit{(Aspect 2)} the energy consumed by with idle resources, e.g., when the resources of an ES are not fully utilized and a portion of them remains unoccupied, represented by: 
\begin{equation}
	E^{\text{S},2}_{j,n} = \Delta t_n  \max (R^{\text{S},\text{In}}_{j,n} - R^{\text{S},\text{Act}}_{j,n} - R^{\text{S},\text{Tra}}_{j,n} + \theta^\text{S}_{j,n}, 0)\eta^{\text{S},\text{Idle}}_{j,n}. 
\end{equation}
Thus, the overall energy consumption of a seller during time frame \( t_n \) is given by: 
\begin{equation}\label{energy2}	
\mathcal{E}^\text{S}_{j,n} = E^{\text{S},1}_{j,n} + E^{\text{S},2}_{j,n}.
\end{equation}

 \vspace{-0.6cm}
\subsection{Modeling of Utility}
We next quantify the utility of different parties (i.e., buyers, sellers, and auctioneer) in our marker of interest.

\noindent$\bullet$ \textbf{Utility and expected utility of buyers:} Let \( \theta_{i,n}^\text{B} \) represent the number of resources defaulted by buyer \( \bm{b}_{i,n} \) during the transaction in time frame \( t_n \). We define the four components of the utility of each buyer as follows: 
\begin{equation}
	U^{\text{B},1}_{i,n} = bid'_{i,n} \min(R^{\text{B},\text{Act}}_{i,n}, R^{\text{B},\text{In}}_{i,n}, R^{\text{B},\text{Tra}}_{i,n}), 
\end{equation}
\begin{equation}
	U^{\text{B},2}_{i,n} = -p^\text{B}_{i,n}(R^{\text{B},\text{Tra}}_{i,n} - \theta_{i,n}^\text{B}), 
\end{equation}
\begin{equation}
	U^{\text{B},3}_{i,n} = -q_{i,n}^\text{B}\theta_{i,n}^\text{B} , 
\end{equation}
\begin{equation}
	U^{\text{B},4}_{i,n} = -\lambda \mathcal{E}^\text{B}_{i,n}, 
\end{equation}
where \( U^{\text{B},1}_{i,n} \) represents the revenue generated by the buyer from providing computing services to its covered vehicles. We assume that the revenue obtained from a resource unit (one RB) is equal to its bid \(\mathsf{bid}'_{i,n} \). Next, \( U^{\text{B},2}_{i,n} \) reflects the net payment made by the buyer for the non-defaulted portion of the contractual resources, accounting for any resources committed under the contracts. Then, \( U^{\text{B},3}_{i,n} \) captures the penalty incurred if the buyer breaks the contract, and \( U^{\text{B},4}_{i,n} \) denotes buyer's energy consumption cost, with \( \mathcal{E}^\text{B}_{i,n} \) representing the energy cost factor. Accordingly, the overall utility of a buyer is given by:
\begin{equation}
	\mathcal{U}^{\text{B}}_{i,n} = U^{\text{B},1}_{i,n} + U^{\text{B},2}_{i,n} + U^{\text{B},3}_{i,n} + U^{\text{B},4}_{i,n}. 
\end{equation}
Given the dynamic nature of resource demands,  obtaining the practical utility of each buyer at \( t_0 \) is challenging. To facilitate the formulation of contracts for future transactions, we thus calculate the expected utility of \(\mathcal{U}^{\text{B}}_{i,n}\) as:
\begin{align}\label{ExpUbuyers}
	\mathbb{E}[\mathcal{U}^{\text{B}}_{i,n}] &= \mathbb{E}[U^{\text{B},1}_{i,n} + U^{\text{B},2}_{i,n} + U^{\text{B},3}_{i,n} + U^{\text{B},4}_{i,n}] \notag\\
	&=\sum_{\kappa = 0}^{R^{\text{B},\text{Tra}}_{i,n}} P_{i,n,\kappa}^\text{B} (\mathsf{bid}'_{i,n} \min(R^{\text{B},\text{Act}}_{i,n}, R^{\text{B},\text{In}}_{i,n}+
	R^{\text{B},\text{Tra}}_{i,n}) 
	\notag\\
	&~~~~~~~~~~~~~~~~~- (R^{\text{B},\text{Tra}}_{i,n} - \kappa) p_{i,n}^\text{B} -  \kappa q_{i,n}^\text{B}-\lambda \mathcal{E}^\text{B}_{i,n}  ), 
\end{align}
where \( P^{\text{B}}_{i,n,\kappa} = \Pr(\theta^\text{B}_{i,n} = \kappa) = \Pr(R^{\text{B},\text{Act}}_{i,n} = R^{\text{B},\text{Tra}}_{i,n} + R^{\text{B},\text{In}}_{i,n} - \kappa) \) represents the probability of needing to default on \(\kappa \) RBs, e.g., \( P^{\text{B}}_{i,n,1} = \Pr(\theta^\text{B}_{i,n} = 1) = \Pr(R^{\text{B},\text{Act}}_{i,n} = R^{\text{B},\text{Tra}}_{i,n} + R^{\text{B},\text{In}}_{i,n} - 1) \). Since actual resource $R^{\text{B},\text{Act}}_{i,n}$ and its probability distribution in \eqref{ExpUbuyers} are difficult to obtain directly, we will later develop a Long Short-Term Memory (LSTM) network to predict future resource demands. This approach enables more accurate and data-driven resource allocation decisions (see Section \ref{sec:Evaluation}).

\noindent$\bullet$ \textbf{Utility and expected utility of sellers:} We assume that sellers prioritize meeting the demands of their contractual buyers (i.e., other ESs) as stipulated in pre-signed agreements. Accordingly, even when a seller’s available resources are insufficient to fully serve its local customers, such as vehicles within its coverage during a given time frame, it remains obligated to honor its contractual commitments. Based on this principle, the utility of a seller is represented by four components:
\begin{equation}
	U_{j,n}^{\text{S},1} = (R_{j,n}^{\text{S},\text{Tra}} - \theta_{j,n}^{\text{S}}) p_{j,n}^{\text{S}},
\end{equation}
\begin{equation}
	U_{j,n}^{\text{S},2} = \theta_{j,n}^{\text{S}} q_{j,n}^{\text{S}},
\end{equation}
\begin{equation}
	U_{j,n}^{\text{S},3} = -\lambda \mathcal{E}_{j,n}^{\text{S}},
\end{equation}
\begin{equation}
	U_{j,n}^{\text{S},4} = v_{j,n}^{\text{S}} \min(R_{j,n}^{\text{S},\text{In}} - R_{j,n}^{\text{S},\text{Tra}} + \theta_{j,n}^{\text{S}}, R_{j,n}^{\text{S},\text{Act}}).
\end{equation}
Here, $v_{j,n}^{\text{S}}$ is the unit profit obtained from its covered vehicles, and $\theta_{j,n}^{\text{S}}$ denotes  the amount of resources for which the seller $\bm{s}_j$ is breached by its contractual buyers during $t_n$. Specifically, $U_{j,n}^{\text{S},1}$ describes the revenue for contributing computing services to buyers; $U_{j,n}^{\text{S},2}$ refers to  the penalty that a seller receives from the buyers who default; $U_{j,n}^{\text{S},3}$ captures the energy cost associated with data analyzing, including both the demand of vehicles covered by this seller, and that from contractual buyers; and $U_{j,n}^{\text{S},4}$ denotes the income obtained from serving its covered vehicles. 
Accordingly, the utility of a seller during $t_n$ is given by:
\begin{equation}\label{sum}
	\mathcal{U}_{j,n}^{\text{S}} = U_{j,n}^{\text{S},1} + U_{j,n}^{\text{S},2} + U_{j,n}^{\text{S},3} + U_{j,n}^{\text{S},4}. 
\end{equation}
The complexity of accurately quantifying a seller's actual resource allocation ($R^{\text{S},\text{Act}}_{j,n}$) and energy consumption ($\mathcal{E}^\text{S}_{j,n}$), and consequently deriving precise operational utility metrics, stems from a key feature of our market design: dynamic M2M buyer-seller mappings with contracts negotiated in advance of actual transactions. In this setup, a seller’s realized utility is subject to post-transaction uncertainties that cannot be fully anticipated at contract formation. Specifically, future buyer defaults and fluctuations in the seller's own resource availability, driven by varying vehicle demand, introduce unpredictability in revenue. These factors make it necessary, yet challenging, to compute the expected utility of each seller. To facilitate calculations, we perform several operations on the intractable elements in $\mathcal{U}^\text{S}_{j,n}$. First, for the estimation of $\theta_{j,n}^{\text{S}}$, we relax the utility expression  $\mathcal{U}_{j,n}^{\text{S}}$, by considering a worst-case scenario that a seller may face in which all buyers breach their contracts. Letting $\ell$ be a non-negative integer, 
we use $P_{j,n,\ell}^{\text{S}} = \Pr(R_{j,n}^{\text{S},\text{Act}} = \ell)$ to represent the probability that a seller has utilized exactly $\ell$ RBs for providing computing services to both its contractual buyers and its own covered vehicles. Based on this, we proceed to outline the step-by-step process for calculating the expected utility of a seller $\bm{s}_{j,n}$. 
In particular, to represent the probability that a seller has utilized exactly $\ell$ RBs for providing computing services to both its contractual buyers and its own covered vehicles. Then, for $U_{j,n}^{\text{S},1}$ and $U_{j,n}^{\text{S},2}$, their theoretical minimum can be reached when all the contractual buyers break their contracts. 
Accordingly, we have the lower bound of the expectation of $U_{j,n}^{\text{S},1}$ and $U_{j,n}^{\text{S},2}$, given by:
\begin{equation}
	\mathbb{E} \left[ U_{j,n}^{\text{S},1} + U_{j,n}^{\text{S},2} \right] \geq R_{j,n}^{\text{S},\text{Tra}} q_{j,n}^{\text{S}}. 
\end{equation}
Next, since  $U_{j,n}^{\text{S},3}$ stands for the energy cost,
 we have its expected value:
\begin{equation}
	\mathbb{E} \left[ U_{j,n}^{\text{S},3} \right] = 
	\sum_{\ell=0}^{R_{j,n}^{\text{S},\text{In}} - R_{j,n}^{\text{S},\text{Tra}}} 
	P_{j,n,\ell}^{\text{S}} \times \left( -\lambda  \mathcal{E}_{j,n}^{\text{S}} \Big| _{R_{j,n}^{\text{S},\text{Act}} = \ell} \right), 
\end{equation}
where $\mathcal{E}_{j,n}^{\text{S}} \Big| _{R_{j,n}^{\text{S},\text{Act}} = \ell} $ represents the condition of $R_{j,n}^{\text{S},\text{Act}} = \ell$ in $\mathcal{E}_{j,n}^{\text{S}}$. 
In addition, in \eqref{sum} $U_{j,n}^{\text{S},4}$ represents the revenue generated by the seller from the vehicles it directly covers.
In the extreme case where all contractual buyers of seller 
$\bm{s}_{j,n}$ default, the seller could reallocate reserved resources to its own customers (i.e., vehicles under its coverage), partially offsetting the loss. However, estimating the expected value of such recovery is highly complex. For analytical simplicity and to maintain a conservative model, we set the component $\theta_{j,n}^\text{S}$ to zero and bound the expected value of  $U_{j,n}^{\text{S},4}$ as:
\begin{equation}
	\mathbb{E} \left[ U_{j,n}^{\text{S},4} \right] \geq v_{j,n}^{\text{S}} \min(R_{j,n}^{\text{S},\text{In}} - R_{j,n}^{\text{S},\text{Tra}}, R_{j,n}^{\text{S},\text{Act}}). 
\end{equation}
Based on the above, we finally get the lower bound of the expected utility of each seller during time frame $t_n$ as follows: 
\begin{equation}\label{expseller}
	\mathbb{E} \left[ \mathcal{U}_{j,n}^{\text{S}} \right] \geq 
	\mathbb{E} \left[ U_{j,n}^{\text{S},1} \right] + 
	\mathbb{E} \left[ U_{j,n}^{\text{S},2} \right] + 
	\mathbb{E} \left[ U_{j,n}^{\text{S},3} \right] + 
	\mathbb{E} \left[ U_{j,n}^{\text{S},4} \right].
\end{equation}
To ensure service delivery while protecting the profits of all ESs, we adopt this lower bound as the minimum acceptable utility for sellers under worst-case conditions and incorporate it as a constraint in our optimization (see Section \ref{3.3}).

\noindent$\bullet$ \textbf{Utility of the auctioneer:} In our market of interest, a trusted auctioneer manages the auction process in two stages. At the first stage (at time $t_0$ prior to actual transactions), it collects buyers' bids and sellers' asks to coordinate the auction and facilitate LAContract signing. At the second stage (start from $t_1$, during the actual services), contracts are implemented under the auctioneer's supervision, ensuring buyers' resource demands are met and sellers receive corresponding revenue.
 
Following contracts, all commissions paid by buyers are fully allocated to corresponding sellers. However, the penalty fees received by sellers may differ from those paid by buyers. In particular, the auctioneer’s revenue is determined by the difference between the penalties collected from buyers and those distributed to sellers. Let $\theta^b_{i,n}$ denote the total amount of defaults attributed to a buyer during time frame $t_n$, and $\theta^\text{S}_{j,n}$ represent the total defaults experienced by seller $\bm {s}_j$ in the same period. Mathematically, the utility of the auctioneer can be represented as:
\begin{equation}
	\mathcal{U}^{a}_n = \sum_{i=1}^{|\mathcal{B}_n|} q_{i,n}^{\text{B}} \theta_{i,n}^{\text{B}} 
	- \sum_{j=1}^{|\mathcal{S}_n|} q_{j,n}^{\text{S}} \theta_{j,n}^{\text{S}}. 
\end{equation}

 \vspace{-2mm}
\subsection{Problem Formulation}\label{3.3}
\textit{The core motivation of this paper is to enable time-efficient, flexible, and energy-aware (green) provisioning of computing services among distributed ESs within a dynamic EdgeIoV environment, where resource demand and supply are inherently uncertain and spatio-temporally variable. By facilitating intelligent and adaptive resource sharing through predictive modeling and pre-negotiated contracts, the proposed framework aims to enhance service responsiveness and reliability.} Ultimately, this not only improves overall system performance but also increases the attractiveness of the EdgeIoV infrastructure to end users, such as vehicles requiring low-latency and dependable computing support. To realize such a goal, we propose a mechanism that encourages ESs, acting as both buyers and sellers, to engage in auctions for pre-signed LAContracts prior to actual service transactions. To support this, we formulate an optimization problem over the observation time horizon across all ESs, with the aim of maximizing their overall expected utility, i.e., social welfare\footnote{Note that the optimization of the auctioneer’s utility is not the primary focus of this work, as the auction design inherently guarantees that the auctioneer will not experience any losses. This is ensured by our auction mechanism, which secures sufficient revenue for the auctioneer to avoid any deficit, as detailed in Section \ref{sec:Methodology Design}.}, as follows:
\begin{align}
	\bm{\mathcal{P}}_0: &\max_{\bm{\mathbb{C}}_n}
	\sum_{n=1}^{|\bm{N}|} \sum_{j=1}^{|{\mathcal{S}_n}|} \sum_{i=1}^{|{\mathcal{B}_n}|} 
	\left( \mathbb{E} \left[ \mathcal{U}_{i,n}^\text{B} \right] + \mathbb{E} \left[ \mathcal{U}_{j,n}^{\text{S}} \right] \right) 
\end{align}
\setcounter{equation}{21}
\vspace{-2mm} 
\begin{subequations}\label{p0}{
\begin{align}
	\text{S.t.} \quad & R_{i,n}^{\text{B},\text{Tra}} \leq R_{i,n}^{\text{S},\text{Est-}}, \quad \forall t_n \in \bm{T}, \bm{s}_{j,n} \in {\mathcal{S}_n} \quad  \label{p0_24a}\\[5pt]
	& R_{j,n}^{\text{S},\text{Tra}} \leq R_{i,n}^{\text{B},\text{Est-}} \quad \forall t_n \in \bm{T}, \bm{b}_{i,n} \in {\mathcal{B}_n}  \quad  \label{p0_24b}\\[5pt]
	& R_{i,j,n}^\text{Con} \geq 0, \quad \forall t_n \in \bm{T}, \forall \text{S}_{j,n} \in \mathcal{S}_n, \forall \text{B}_{i,n} \in \mathcal{B}_n \quad\label{p0_24c} 
\end{align}}
\end{subequations}

\noindent In $\bm{\mathcal{P}}_0$,  constraints \eqref{p0_24a} and \eqref{p0_24b} impose resource feasibility by ensuring that the cumulative resources allocated by each seller do not surpass its estimated supply, and that each buyer’s reserved resources do not exceed its anticipated demand. Additionally, constraint \eqref{p0_24c} enforces \textit{non-negativity}, ensuring that all resource allocations specified in the LAContracts are valid.

It is easy to verify that $\bm{\mathcal{P}}_0$ represents an NP-hard  mixed-integer nonlinear programming (MINLP) problem \cite{NPhard}, calling for addressing both contract terms (e.g., resource prices, payments, penalties that are continuous values), and proper M2M mapping in each time frame (discrete/binary in nature). Besides, the dynamic and uncertain environment with fluctuant resource demand/supply introduces additional complexity to the problem, such as the need to determine the role (buyer or seller) of each ES across the entire time horizon, from $t_1$ to $t_N$. 

To tackle this complex problem, in section \ref{sec:Methodology Design}, \textit{we decouple it into two subproblems}. The former subproblem focuses on predicting resource demand and supply within each discrete time frame, aiming to obtain tractable expected utility formulations. Based on these predictions, the second subproblem introduces a double auction mechanism, which determines the optimal buyer-seller matchings and pricing rules for LAContracts, thereby enabling the establishment of pre-signed contracts among ESs for each individual time frame. Collectively, the solutions to these two subproblems 
 address problem $\bm{\mathcal{P}}_0$. We next present a set of properties for our market of interest that our solutions aim to satisfy.
 \vspace{-0.2cm}
\subsection{Desirable Properties}\label{3.4}
\textit{A distinctive feature of our work is that the double auction process is conducted prior to actual resource transactions, specifically designed to establish LAContracts for future time frames.} This innovation introduces fundamental differences in the auction properties compared to conventional auction models in existing literature. In the following, key design properties of our market of interest are introduced. 
\begin{definition}[Individual Rationality in Our Market]\label{def:definition1}
	An auction is said to be \emph{individually rational} in our context if each participant (buyer or seller) satisfies the following conditions:
	\begin{enumerate}[(i)]
		\item \textbf{Non-negative expected profit per time frame:} The expected utility of both buyers and sellers for each time frame is non-negative.
		\item \textbf{Non-negative cumulative return:} After all contracted transactions are completed, the final cumulative utility for each participant is non-negative.
		\item \textbf{Superiority over non-participation:} The total income of any participant is no less than what they would earn by not participating in the auction.
	\end{enumerate}
\end{definition}
\begin{remark} (Our Unique Angle of View Regarding Individual rationality). \textit{In conventional auctions, individual rationality requires that both buyers and sellers achieve non-negative utility in each transaction. In our setting, this notion is extended in several ways. First, ESs may act as sellers for their own end-users (i.e., vehicles under their coverage), prioritizing contractual buyers during transactions. This prioritization does not necessarily incur a loss, as LAContract amounts are estimated from historical data; however, if serving their own end-users is more profitable, participation may be less attractive. Second, ESs acting as buyers may face losses due to penalties for contract breaches. Thus, an effective market mechanism must incentivize participation by offering reasonably priced services for efficient resource integration. Finally, to sustain confidence, especially when signing multiple contracts initially, our redefined individual rationality considers both short-term transaction profitability and long-term cumulative returns, ensuring participation is preferable to non-participation.}
\end{remark}

\begin{definition}[Budget Balance]
\textit{Budget balance} implies that the total charges from buyers does not exceed the total payments to the sellers, namely, the utility of the auctioneer must be non-negative.
\end{definition}
\begin{definition}[Truthfulness or Incentive Compatibility]
A truthful auction guarantees that all buyers/sellers will report their bids/asks equal to their true valuation during the auction for  LAContracts, since misreporting will not result in improved utility. 
\end{definition}
 In the following sections, we first present the detailed solution for problem $\bm{\mathcal{P}}_0$, and subsequently prove that our proposed methodology satisfies the above design properties.

\vspace{-0.1cm}
\section{Solution Design and Property Analysis}\label{sec:Methodology Design}
As previously discussed, accurately estimating future resource usage based on historical data plays a critical role in facilitating the formation of LAContracts among ESs. To address this, we decompose  $\bm{\mathcal{P}}_0$ into two subproblems:

\noindent
$\bullet$ The first subproblem, denoted as $\bm{\mathcal{P}}_1$, is formulated as a \textit{regression task} for predicting future usage of ESs.

\noindent
$\bullet$ The second subproblem, represented by $\bm{\mathcal{P}}_2$, is modeled as an \textit{optimization problem} under a double auction framework within each time frame, aiming to maximize the overall expected utilities (i.e., social welfare) of participating ESs.

\begin{figure}[h]
\vspace{.7mm}
	\centering
	\includegraphics[width=\linewidth]{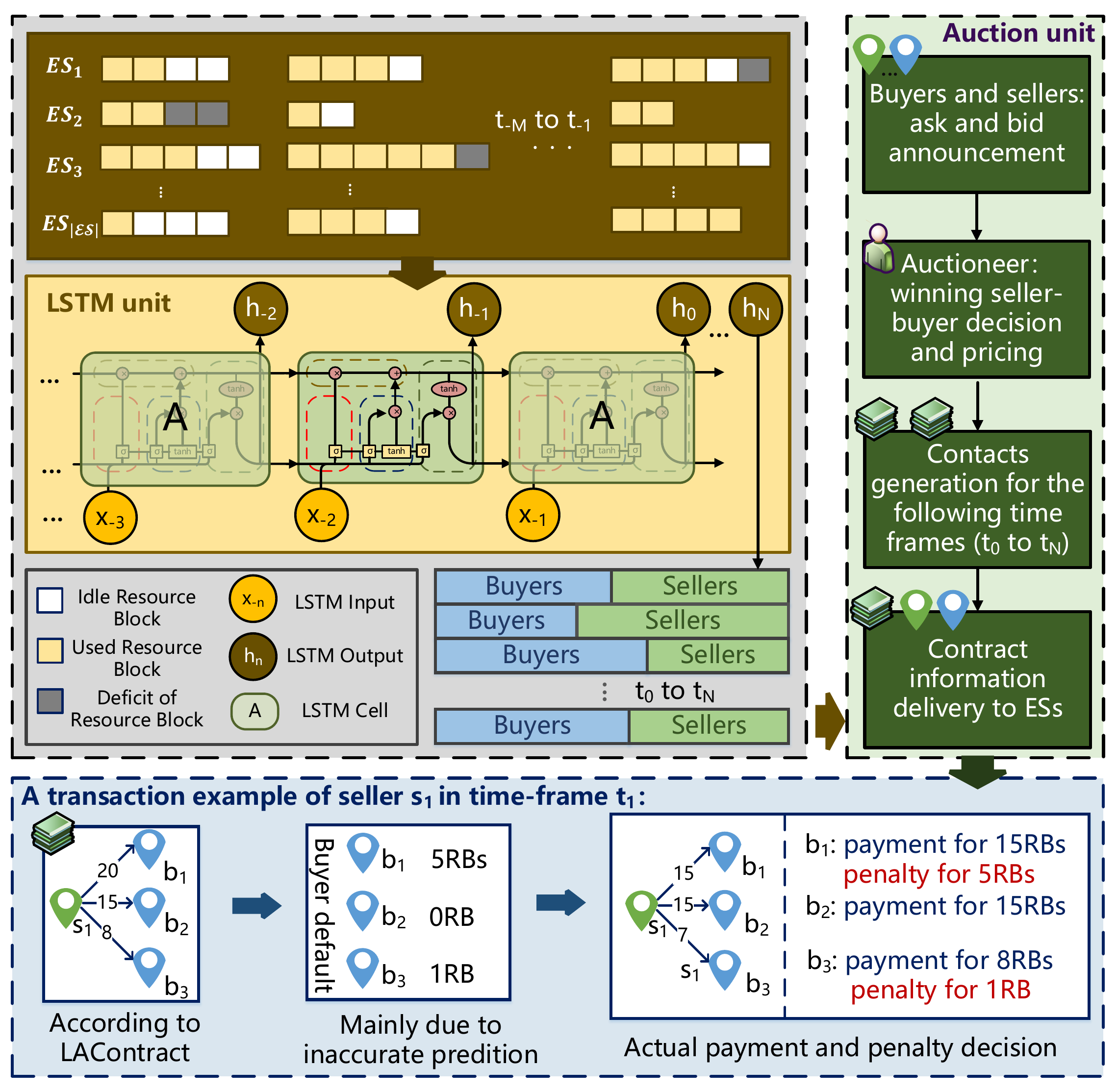}
    \vspace{-6.5mm}
	\caption{Schematic diagram of LSTM-based resource usage prediction and role determination (the gray box), the pre-double auction process (the green box), and contract implementation (the blue box).}
	\label{fig:LSTM}
	\vspace{-0.6cm}
\end{figure}

In a nutshell, for the first subproblem, we employ an LSTM-based prediction model to estimate the future resource usage of each ES. Based on these estimations, each ES can determine its preferred role (i.e., buyer or seller) in each upcoming time frame. Subsequently, for the second subproblem, we design a mechanism to determine the winning buyer-seller pairings and corresponding pricing rules, thereby facilitating the signing of LAContracts among ESs with differing roles in each time frame. An overview of this two-stage paradigm is illustrated in Fig. \ref{fig:LSTM}.
\vspace{-5mm}
\FloatBarrier

\vspace{0.01cm}
\subsection{Prediction of Future Resource Usage for ESs and Their Role Determination}\label{4.1}
We use the symbol ${R}_{k,n}^{\text{Est}}$ to represent the predicted ES resource usage during time frame $t_n$, where $\forall ES_k \in \mathcal{ES}$, which helps determine the roles of ESs in each transaction, as well as the expected utilities of buyers an sellers embedded in $\bm{\mathcal {P}}_0$. 
To effectively capture the dynamic nature of the EdgeIoV environment, the first subproblem is formulated as a regression task, with the objective of minimizing the mean squared error (MSE) in estimating future resource usage of all ESs over the entire time horizon, as given by:
\begin{align}\label{p1}
	\bm{\mathcal{P}}_1: \quad  \min_{{R}_{k,n}^{\text{Est}} }\frac{1}{|\mathcal{ES}|} \sum_{k=1}^{|\mathcal{ES}|}\frac{1}{N} \sum_{n=1}^{N} 
	&({R}_{k,n}^{\text{Est}}-R_{k,n}^{\text{Act}})^2.
	\quad
\end{align} 
In $\bm{\mathcal{P}}_1$, ${R}_{k,n}^{\text{Est}}$ and ${R}_{k,n}^{\text{Act}}$ are crucial parameters in the previous utility calculations (e.g., ${R}_{k,n}^{\text{Est-}}$, ${R}_{k,n}^{\text{Tra}}$). For buyers, prediction accuracy directly affects the penalties they may incur, while for sellers, it impacts the profits they can realize.
 Since the prediction of future resource usage is assumed to be independent across different ESs, this problem can further be decoupled\footnote{Although ESs are geographically distributed, the vehicle traffic they serve is not entirely independent. Specifically, when two ESs are located in close proximity, vehicles within the coverage area of one ES during a given time slot may migrate to the neighboring ES in subsequent slots, resulting in an implicit coupling of their service demands. However, the primary focus of this paper is to establish a novel market-based framework for looking $N$-steps ahead of future resource trading among ESs. Thus, to maintain conceptual clarity and emphasize the core contributions, we neglect these lower-level mobility dynamics in the current model.} (i.e., each ES minimizes the MSE between its own predicted and actual resource usage). Accordingly, the  $\bm{\mathcal{P}}_1$ can be reformulated from the perspective of individual ES as a series of subproblems, each formulated as follows:
\vspace{-.75mm}
\begin{align}\label{p1k}
	\bm{\mathcal{P}}_{1,k}: \min_{{R}_{k,n}^{\text{Est}}} 
	\frac{1}{N}\sum_{n=1}^{N} 
	&({R}_{k,n}^{\text{Est}}-R_{k,n}^{\text{Act}})^2.
	\quad
\end{align} 
To address \eqref{p1k}, we adopt the LSTM network as an effective tool for estimating the resource usage of each ES in every time frame. Prior studies \cite{LSTM} have demonstrated the superior performance of LSTM networks in time series forecasting tasks, particularly due to their ability to capture long-term dependencies and mitigate the vanishing gradient problem through specialized gating mechanisms. In our study, we utilize hourly data with moderately long input sequences (e.g., the past 168 hours) to forecast future resource usage (e.g., over the next 24-hour period), with details given in Section \ref{sec:Evaluation}.  
For each $ES_k \in \mathcal{ES}$, we denote its historical data as  $R^{\text{Act}}_{k,-\overline{N}}$ to $R^{\text{Act}}_{k,-1}$ (assuming that we have historical information about $\overline{N}$ time frames) as the input of our LSTM unit (see Fig. \ref{fig:LSTM}). As the model output,  the resource usage for the next $N$ time frames ${R}_{k,1}^{\text{Est}}$ to ${R}_{k,N}^{\text{Est}}$ can be obtained. Based on which, each ES determines its role: for example, if ${R}_{k,n}^{\text{Est}}-R_{k,n}^{\text{In}} > 0$, it serves as a buyer
and we have $R_{i,n}^{\text{B},\text{Est-}}$ calculated by  ${R}_{i,n}^{\text{Est}}-R_{i,n}^{\text{In}}$. In addition, each ES can determine the quantity of resources to purchase or offer based on the discrepancy between its estimated demand  ${R}_{k,n}^{\text{Est}}$ and resource availability  $R_{k,n}^{\text{In}}$. 
These insights further lay the groundwork for the double auction mechanism designed for establishing LAContracts, as we describe next.

\vspace{-0.2cm}
\subsection{Design of Pre-Double Auction for LAContracts}
Building on the intelligent predictive framework inherently captures temporal dependencies across consecutive time frames, 
the second subproblem is formulated as an optimization problem \(\bm{\mathcal{P}}_2\) given in \eqref{p2}, aiming to maximize the overall expected utility of both buyers and sellers within each individual upcoming time frame $t_n$:
\begin{align}
	\bm{\mathcal{P}}_2: \quad & \max_{\bm{\mathbb{C}}_n} \sum_{j=1}^{|\mathcal{S}_n|} \sum_{i=1}^{|\mathcal{B}_n|}
	\left( \mathbb{E} \left[ \mathcal{U}_{i,n}^{\text{B}} \right] + \mathbb{E} \left[ \mathcal{U}_{j,n}^{\text{S}} \right] \right)
	\quad\label{p2}
\end{align}
\setcounter{equation}{24}
\vspace{-2.5mm} 
\begin{subequations}\label{p0}{
		\begin{align}
			\text{S.t.} \quad&  
			R_{i,n}^{\text{B},\text{Tra}} \leq R_{j,n}^{\text{S},\text{Est-}}, \quad \forall \text{S}_{j,n} \in \mathcal{S}_n \quad \\[5pt]
			&R_{j,n}^{\text{S},\text{Tra}} \leq R_{i,n}^{\text{B},\text{Est-}}, \quad \forall \bm{b}_{i,n} \in \mathcal{B}_n \quad \\[5pt]
			&R_{i,j,n}^{\text{Con}} \geq 0, \quad \forall \bm{s}_{j,n} \in \mathcal{S}_n, \forall \bm{b}_{i,n} \in \mathcal{B}_n \quad 
	\end{align}}
\end{subequations}

\noindent We address this optimization problem within a double auction framework, structured into two sequential phases.
\textit{Phase 1} involves the determination of optimal buyer-seller matchings, i.e., selecting the set of buyer-seller pairs that are eligible to establish LAContracts based on their submitted bids and asks. \textit{Phase 2} focuses on monetary elements determination, which specifies the contractual terms for each matched pair. This includes determining the payment obligations of buyers, the corresponding revenues for sellers, and the penalty fees applied in the event of contract breaches by buyers. The pseudo-code of these phases  are presented in Algs 1 and 2. In the following, we provide a detailed exposition of the design rationale, decision logic, and economic properties underpinning our  auction mechanism.
\vspace{-2mm}

\begin{algorithm}[h]
	\footnotesize
	\caption{Winning buyer-seller pair determination for time frame $t_n$ }\label{algorithm1}
	\SetKwInOut{Input}{Input}\SetKwInOut{Output}{Output}
	\Input{$\mathcal{S}_n$, $\mathcal{B}_n$, $\bm{A}_n$, $\bm{B}_n$,  $\bm{D}_n$}
	\Output{$\bm{\mathbb{M}}_n$}
	Initialization: $\bm{M}_{i,j,n} \gets \{0, 0, \dots, 0\}, \forall i,j$
	
	Sort the asks $\bm{A_n}$ increasing order, i.e.,:
	\(\bm{A'_n} = \left[(\mathsf{ask}_{1,n}, R_{1,n}^{\text{S},\text{Est-}}),\right. \)
		\(\left.(\mathsf{ask}_{2,n},R_{2,n}^{\text{S},\text{Est-}}),\ldots,(\mathsf{ask}_{j,n},R_{j,n}^{\text{S},\text{Est-}}),\ldots,(\mathsf{ask}_{|\mathcal{S}_n|,n},R_{|\mathcal{S}_n|,n}^{\text{S},\text{Est-}})\right],\)
		\(0\leq\mathsf{ask}_{1,n}\leq\mathsf{ask}_{2,n}\leq\ldots\leq\mathsf{ask}_{j,n}\leq\ldots\leq\mathsf{ask}_{|\mathcal{S}_n|,n}\)
		
	\For{$j=1$ to $|\mathcal{S}_n|$}{
		
		\For{$i=1$ to $|\mathcal{B}_n|$}{
		\(\textbf{find } \mathsf{bid}_{i*,j,n} = \max \left\{ \mathsf{bid}_{1,j,n}, \mathsf{bid}_{2,j,n}, \dots, \mathsf{bid}_{i,j,n}, \dots, \mathsf{bid}_{|\mathcal{B}n|,j,n} \right\}\) 
		
		\If{\( \!\quad\! \mathsf{bid}_{i*,j,n}\! > \! \mathsf{ask}_{j,n} \!\quad\! \textbf{and}\! \quad \! R_{i,n}^{\textnormal{B},\textnormal{Est-}}\!\neq\! 0 \quad \! \textbf{and}\! \quad \! R_{j,n}^{\textnormal{S},\textnormal{Est-}}\!\neq 0 \)}{
		
		Record all non-winning bids satisfying ask:
		\( \quad \quad \quad
		\mathsf{Nbid}_{i,j,n}\!\! \gets\!\! \left\{ \mathsf{bid}_{x,j,n} \mid \mathsf{ask}_{j,n} < \mathsf{bid}_{x,j,n} < \mathsf{bid}_{i,j,n} \right\}\)

		Record both trading parties and the trade quantity: \(R_{i,j,n}^{\text{Con}} = \min \left( R^{b,\text{Est-}}_{i,n} , R^{\text{S},\text{Est-}}_{j,n} \right)\), 
		\(c_{i,j,n}  =\mathsf{bid}'_{i,j,n} - \mathsf{bid}_{i,n} \), 
		\( \bm{M}_{i,j,n} \gets \{ \beta_{j,n}^\text{S}, \beta_{i,n}^\text{B}, R^\text{Con}_{i,j,n}, c_{i,j,n}, \mathsf{Nbid}_{i,j,n} \} \)
		
		Update the demand list: 
		\(
		R_{i,n}^{\text{B},\text{Est-}} = R_{i,n}^{\text{B},\text{Est-}} - \min \left( R_{i,n}^{\text{B},\text{Est-}}, R_{j,n}^{\text{S},\text{Est-}} \right)
		\)
		\(
		R_{j,n}^{\text{S},\text{Est-}} = R_{j,n}^{\text{S},\text{Est-}} - \min \left( R_{i,n}^{\text{B},\text{Est-}}, R_{j,n}^{\text{S},\text{Est-}} \right)
		\)
		
	}
	
		}
	}
	{\bf{return}} $\bm{\mathbb{M}}_n$
\end{algorithm}	

\noindent $\bullet$ \textit{Phase 1 (Alg. 1)}. Under given ESs set with specific roles, to determine winning seller-buyer pairs, we first introduce key input elements: sets of buyers and sellers taking part in the auction to sign LAContracts for future time frame $t_n$, namely, $\mathcal{S}_n$ and $\mathcal{B}_n$, with information about resource supply and demand (as predicted); the ask matrix $\bm{A}_n$ of sellers and the bid matrix $\bm{B}_n$ of buyers; the  distance matrix $\bm{D}_n$ where $[\bm{D}_n]_{i,j} = [d_{i,j,n}]$.   Our goal during this step is to obtain the optimal matching set, 
with \(\bm{\mathbb{M}}_n = \left[\bm{M}_{i,j,n}\right]_{{i\in\{1,...,|\bm{\mathcal{B}_n}|\}},{j\in\{1,...,|\bm{\mathcal{S}_n}|\}}}\) recording matchings between sellers and buyers, 
where \( \bm{M}_{i,j,n} \)
holds the matching details between buyer \( \bm{b}_{i,n} \) and seller \( \bm{s}_{j,n} \) during \( t_n \), defined as \( \bm{M}_{i,j,n} = \{ \beta_{j,n}^\text{S}, \beta_{i,n}^\text{B}, R^\text{Con}_{i,j,n}, c_{i,j,n}, \mathsf{Nbid}_{i,j,n} \} \). Here, \(\beta_{j,n}^\text{S}, \beta_{i,n}^\text{B}\) are the identifiers of corresponding participants, respectively; 
\( R^\text{Con}_{i,j,n} \) represents the amount of trading resources stipulated in the contract; \(c_{i,j,n} = \mathsf{bid}'_{i,j,n} - \mathsf{bid}_{i,n} \) quantifies the reduction in the buyer’s bid caused by transmission overhead (per unit data/RB), i.e., the amount by which the seller’s received payment decreases due to the transmission distance. For example, if a single resource block RB can process 500 KB of data during time frame  $t_1$, then \(c_{2,3,1}\) denotes the cost that buyer $\bm{b}_{2,1}$ incurs to transmit this 500 KB to seller $\bm{s}_{3,1}$ within $t_1$.

\noindent \textbf{Step 1 (Ask Ordering).}
To ensure fairness in resource allocation, sellers are first sorted in ascending order of their asking prices (line 2, Alg. 1). This prioritizes those offering lower prices, which is consistent with conventional market trading principles.

\noindent\textbf{Step 2 (Bid Screening).}
For each seller in this ordered list, the buyer with the highest bid that meets or exceeds the seller’s asking price is selected as the winning buyer (lines 5–6, Alg. 1). Meanwhile, all other non-winning bids that also satisfy the asking price are recorded into the set $\mathsf{Nbid}_{i,j,n}$ (line 7, Alg. 1). This information is later used for determining transaction prices (see Alg. 2).

\noindent \textbf{Step 3 (Information Update).}
Once a match is made, the trading quantity for the current time frame is recorded, along with the agreed prices and transmission costs. To enable flexible M2M matching, a seller is not required to fully satisfy a buyer’s demand in one transaction. If only part of the demand is fulfilled, the remaining demand is updated (line 9, Alg. 1) and carried forward to subsequent rounds of matching (still associated with the original bid price).

\noindent \textbf{Step 4 (Iteration and Termination).}
The matching process continues iteratively: after one seller is processed, the next seller in the ascending order is considered. This procedure repeats until all sellers have either been matched with eligible buyers or no further buyers remain.

\noindent \textbf{Step 5 (Result Formation).}
At the conclusion of the process, we obtain the set of winning seller–buyer pairs, denoted by $\bm{\mathbb{M}}_n$. Each entry in $\bm{\mathbb{M}}_n$ contains the seller’s asking price, the buyer’s bidding price, the agreed trading quantity, and the list of competing non-winning bids. This information is preserved to support the subsequent pricing mechanism.

To ensure that both buyers and sellers achieve reasonable and incentivizing utilities, it is critical to establish appropriate pricing rules for computing services, as well as compensation terms in the event of contract breaches. Accordingly, in Phase 2, we design a contract pricing scheme that determines monetary contract terms, such as the buyer’s payment and the seller’s compensation in case of buyer default.
\begin{algorithm}[t!]
	\footnotesize
	\caption{LAContract term determination}\label{algorithm2}
	\SetKwInOut{Input}{Input}\SetKwInOut{Output}{Output}
	\Input{$\bm{\mathbb{M}}_n$, $\mathcal{S}_n$, $\mathcal{B}_n$}
	\Output{$\bm{\mathbb{C}}_n$}
	Initialization: $\bm{C}_{i,j,n}\gets\{ 0,0,\dots,0\}, \forall i,j $
	 
	\textbf{\# Step 1: Price determination}
	
	\For{ each $\bm{M}_{i,j,n} \in \bm{\mathbb{M}}_n$ }{
		\If{$R^\textnormal{Con}_{i,j,n} \neq 0$}{
			
			$p_{i,j,n} \gets \text{Avg}(\mathsf{Nbid}_{i,j,n})$\;
			\tcp{Compute price:The average number of buyers who bid more than the seller asked but did not win the auction }
			}
	}
	\For{$j=1$ to $|\mathcal{S}_n|$}{
		$p^\text{S}_{j,n} =\frac{1}{R^{\text{S},\text{Tra}}_{i,j}}\sum_{i=1}^{|\mathcal{B}_n|} p_{i,j,n}R^\text{Con}_{i,j,n}$\;
	}
	\For{$i=1$ to $|\mathcal{B}_n|$}{
		$p^\text{B}_{i,n} = \frac{1}{R^{\text{B},\text{Tra}}_{i,j}}\sum_{j=1}^{|\mathcal{S}_n|}(p_{i,j,n} + c_{i,j,n})R^\text{Con}_{i,j,n}$\;
	}

	\textbf{\# Step 2: Penalty determination}
	
	\For{$j=1$ to $|\mathcal{B}_n|$}{	
		\For{$i=1$ to $|\mathcal{S}_n|$}{
			\If{$R_{i,j,n}^\textnormal{Con} \neq 0$}{
				$q^\text{B}_{i,n} = \max(q^\text{B}_{i,n} , q^\text{S}_{j,n})$ 
			}
		}
	}
	
	\textbf{\# Step 3: Update contracts' information}
	
	\For{each $\bm{M}_{i,j,n} \in \bm{\mathbb{M}}_n$ and each $\bm{C}_{i,j,n} \in \bm{\mathbb{C}}_n$}{
			\If{$R_{i,j,n}^\textnormal{Con} \neq 0$}{
				$\bm{C}_{i,j,n}\gets\{ \beta_{j,n}^\text{S}, \beta_{i,n}^\text{B}, R^\text{Con}_{i,j,n}, c_{i,j,n} , p^\text{B}_{i,n}, p^\text{S}_{j,n}, q_{i,n}^\text{B},q_{j,n}^\text{S}\}$
			}
}
	{\bf{return}} $\bm{\mathbb{C}}_n$
\end{algorithm}	

\noindent $\bullet$ \textit{Phase 2 (Alg. 2)}. In this step, relying on the matching records $\bm{\mathbb{M}}_n$ generated by Alg. 1 as input, we aim to obtain the proper contract set $\bm{\mathbb{C}}_n$ for time frame $t_n$. In particular, \( \bm{C}_{i,j,n} = \{ \beta_{j,n}^\text{S}, \beta_{i,n}^\text{B}, R^\text{Con}_{i,j,n}, c_{i,j,n} , p^\text{B}_{i,n}, p^\text{S}_{j,n}, q_{i,n}^\text{B},q_{j,n}^\text{S}\} \) represents the contract (if a match exists) between buyer $\bm{b}_i$ and seller $\bm{s}_j$, where parameters $\beta_{j,n}^\text{S}, \beta_{i,n}^\text{B}, R^\text{Con}_{i,j,n}, c_{i,j,n}$ are based on elements in  $\bm{M}_{i,j,n}$ (line 19, Alg. 2). Meanwhile, each contract in list $\bm{\mathbb{C}}_n$ contains the unit payment (per RB\footnote{The final price consists of $p^\text{S}_{j,n}$ and $p^\text{B}_{i,n}$, representing the seller's receipt and the buyer's unit payment per RB, respectively.} unit) and the penalty incurred due to contract breaches. Obtaining monetary LAContract terms involves three key steps. 

\noindent
 \textbf{Step 1} aims to determine proper prices, including the buyer's unit payment and the seller's unit revenue per RB. To ensure the truthfulness of our auction mechanism, the payment per RB $p_{i,j,n}$ for each winning buyer is set to the average bid of those buyers who met the seller’s asking price but did not win the auction (lines 3-5, Alg. 2). This pricing rule incentivizes truthful bidding while preserving social welfare.
Subsequently, a uniform unit price is determined for both sides: for each seller, $p^\text{S}_{j,n}$ is calculated as the average effective price received from matched buyers (line 6, Alg. 2); for each buyer, $p^\text{B}_{i,n}$ is computed by averaging all payments across matched sellers, while incorporating the data transmission cost $c_{i,j,n}$ (line 8, Alg. 2). As noted in \eqref{bid}, this cost component ensures a realistic total payment that reflects actual resource acquisition and usage overhead in an EdgeIoV.

\noindent
\textbf{Step 2} considers one of our key ideas on penalty design, which enhances system fault tolerance and economic fairness.  In dynamic settings, buyers may fail to execute pre-signed contracts (i.e., not purchase reserved resources at $t_n$).  To allow flexible buyer behavior while protecting seller interests, our framework permits buyer-side default, provided that penalty fees are paid.  These fees are meant to compensate sellers for resource reservation losses and are calculated with the following design:
First, the buyer-side penalty $q^\text{B}_{i,n}$ is set to the maximum penalty among all corresponding sellers (lines 10-14, Alg. 2).  This standardization simplifies buyers' expected utility estimation and ensures fairness.
While the standardized penalty is paid by the buyer, each seller still receives the original penalty $q^\text{S}_{j,n}$ they declared, typically reflecting standby energy costs.  The difference is retained by the platform as a service fee.  This design balances the needs of both sides: sellers are compensated adequately, buyers face predictable consequences, and the system avoids resource underutilization.  Moreover, in the event of buyer default, the released resources can be reallocated to serve the seller's own vehicular users during the same time frame, improving system flexibility and resilience.

\noindent
\textbf{Step 3} focuses on updating and finalizing contract information (lines 15-20,  Alg. 2). For each valid matching record $\bm{M}_{i,j,n}$, we generate a monetary contract $\bm{C}_{i,j,n}$ that contains the complete transaction terms: resource allocation ($R^\text{Con}_{i,j,n}$), buyer/seller weights ($\beta^\text{B}_{i,n}, \beta^\text{S}_{j,n}$), unit payment/revenue ($p^\text{B}_{i,n}, p^\text{S}_{j,n}$), transmission cost $c_{i,j,n}$, and penalties ($q^\text{B}_{i,n}, q^\text{S}_{j,n}$). If no match is found between buyer $\bm{b}_i$ and seller $\bm{s}_j$, we set all the elements in  $\bm{C}_{i,j,n}$ as zero. These zero-filled entries serve as flags for invalid (unmatched) contracts.

\vspace{-2mm}
\subsection{Analysis of Economic Properties}

Our work offers a novel approach by incorporating future resource usage predictions and an advance double auction mechanism that looks $N$ steps ahead for resource trading. Consequently, these properties should hold not only at the contract signing phase (i.e., at $t_0$) but also remain valid throughout the entire time horizon, encompassing all transactions from a long-term perspective. We carry out a set of analysis for demonstrating that our methodology satisfies key economic properties (defined in Section \ref{3.4}). 

\theoremstyle{plain}
\newtheorem{theorem}{Theorem} 
\newtheorem{lemma}{Lemma}[theorem] 
\begin{theorem}
	The proposed methodology ensures individual rationality for all the ESs.
\end{theorem}
\begin{lemma}\label{lemma:1.1}
	The proposed double auction during $t_0$ holds individual rationality for all ESs acting as either buyers or sellers.
\end{lemma}
\begin{lemma}\label{thm:theorem2}
	Implementing the pre-signed LAContracts during practical transactions (i.e., from $t_1$ to $t_N$ ) guarantees non-negative profits for ESs.
\end{lemma}
\begin{lemma}\label{lemma:1.3}
	Implementing the pre-signed LAContracts during practical transactions (i.e., from $t_1$ to $t_N$) can hold the profit of ESs higher than without participating in this market.
\end{lemma}
\begin{theorem}
	Our methodology is budget balance.
\end{theorem}
\begin{lemma}\label{lemma2.1}
	Our methodology is budget balance when there is no contract breaking events.
\end{lemma}
\begin{lemma}\label{lemma2.2}
	Our methodology is budget balance when some buyers break their contracts.
\end{lemma}
\begin{theorem}
	Our methodology can support truthfulness for ESs.
\end{theorem}
\begin{lemma}\label{lem:3.1}
	Buyers in our considered market are truthful. 
\end{lemma}
\begin{lemma}\label{lem:3.2}
	Sellers in our considered market are truthful.
\end{lemma}
 To maintain coherence and due to space constraints, the respective proofs are moved to the Appendix. 

\vspace{-2mm}
\section{Evaluation}\label{sec:Evaluation}
To assess the effectiveness of our proposed methodology, hereafter referred to as \textit{N-step-LATrade} for brevity, we conduct a series of experiments using MATLAB R2023b and python 3.8. The experiments consists of two complementary components: the first leverages a \textit{real-world dataset} to demonstrate the practical performance of $N$-step-LATrade under realistic traffic conditions, while the second is based on \textit{synthetic numerical datasets}, designed to test a range of problem scales and validate the scalability and generalizability of approach across diverse settings.

To substantiate the superiority of our method, we compare its performance against the following benchmarks:

\noindent
$\bullet$ \textit{ConAuction}: This method employs conventional double auction for edge resource sharing among multiple ESs. It operates in an online manner without incorporating resource demand/supply prediction, as inspired by the double auction approach used in \cite{dynamic2}.

\noindent
$\bullet$ \textit{DistaTrade}: Similar to our methodology, this method consists of two stages. However, it adopts a greedy-based strategy, wherein, during LAContract design, sellers select the nearest buyers who meet their asking prices for each transaction. This is inspired by the approach in \cite{houyi}.

\noindent
$\bullet$ \textit{RanTrade}: This method performs online resource sharing without resource usgae prediction. Sellers and buyers are randomly matched in each transaction, without consideration of price or proximity, which is inspired by \cite{random}.

\noindent
$\bullet$ \textit{NoTrade}: This method assumes that ESs do not share resources among each other. Instead, each ES exclusively allocates resources to its own users (vehicles).

Also, to conduct a multi-dimensional evaluation, we consider the following performance metrics:

\noindent
$\bullet$ \textit{Social welfare}: In each time frame $t_n$, the social welfare refers to the aggregate utility  of ESs, calculated by  $\sum_{j=1}^{|{\mathcal{S}_n}|} \sum_{i=1}^{|{\mathcal{B}_n}|} 
\left( \mathcal{U}_{i,n}^{\text{B}}  +  \mathcal{U}_{j,n}^{\text{S}} \right) $.

\noindent
$\bullet$ \textit{Time efficiency}: This metric serves as a critical consideration in our study, capturing the efficiency of service delivery. It is evaluated based on the execution time, specifically, the running time of the MATLAB program responsible for performing buyer-seller mapping and service delivery. This measure reflects the practical responsiveness of a resource trading framework and its suitability for real-time decision-making in dynamic EdgeIoV environments.


\noindent
$\bullet$ \textit{Resource utilization}: This metric is defined as the ratio of computing resources utilized for task execution to the total available resources. It reflects the system’s efficiency in minimizing resource waste and indicates how effectively a resource trading framework leverages available ES capacity to meet service demands in the EdgeIoV environment.

\noindent
$\bullet$ \textit{Energy efficiency}: This metric represents another key indicator within our market framework, defined as the ratio of energy consumed for task processing to the total energy consumption. It reflects the efficiency of energy utilization and serves as a measure of energy waste. 

%
%
%

	\vspace{-3.5mm}
\subsection{Experiments on Real-World UTD19 Dataset}\label{5.1}
\begin{figure}[b]
	\vspace{-5.4mm}
	\centering
	\subfigure[]{
		\includegraphics[trim=0.2cm 0cm 2cm 0cm, clip,
		width=0.31\columnwidth]{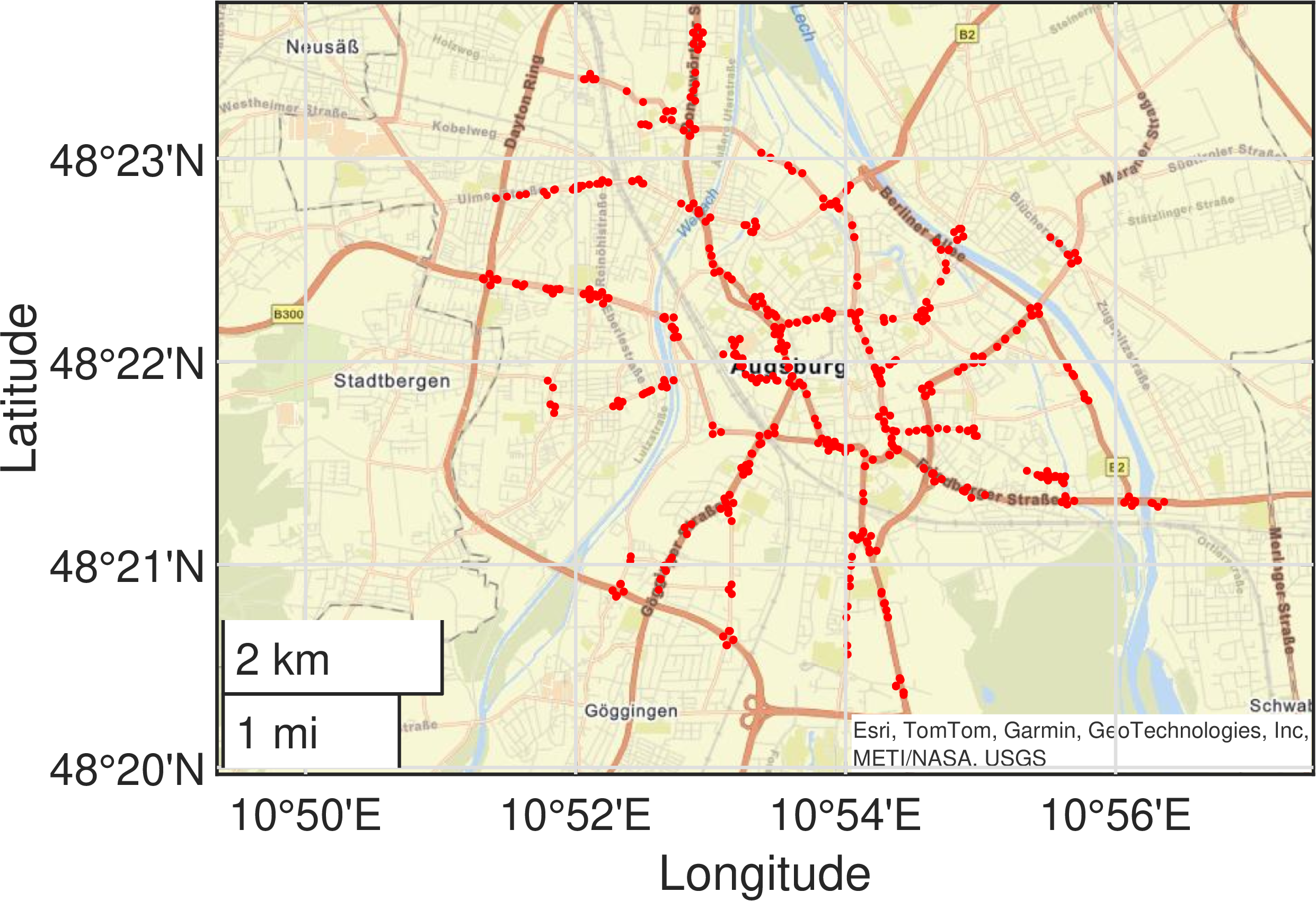}
		\label{fig:detmap}}
	\subfigure[]{
		\includegraphics[trim=0.2cm 0cm 2cm 0cm, clip, width=0.3\columnwidth]{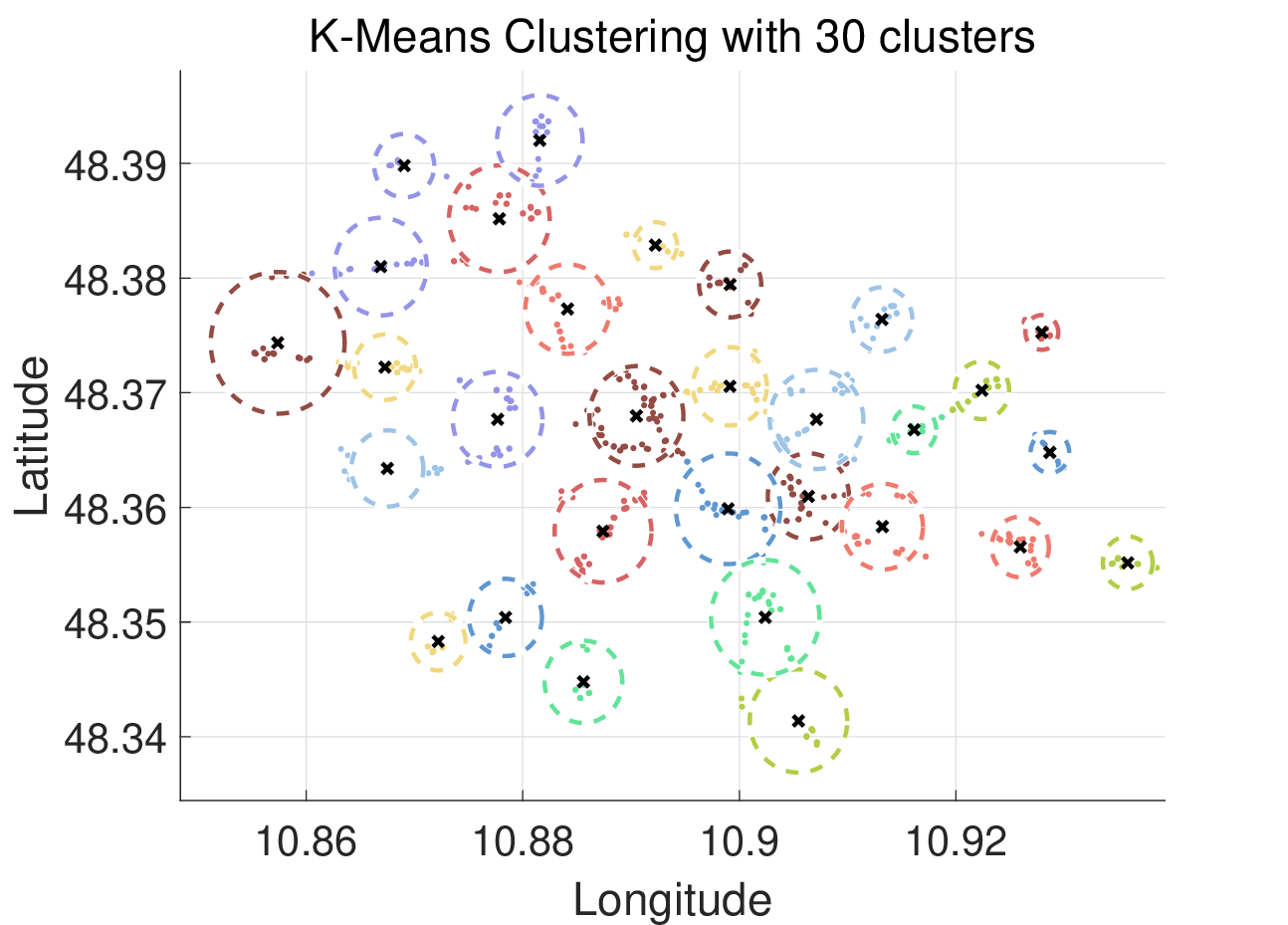}
		\label{fig:k-means30}       
	}
	\subfigure[]{
		\includegraphics[trim=0.2cm 0cm 2cm 0cm, clip, width=0.3\columnwidth]{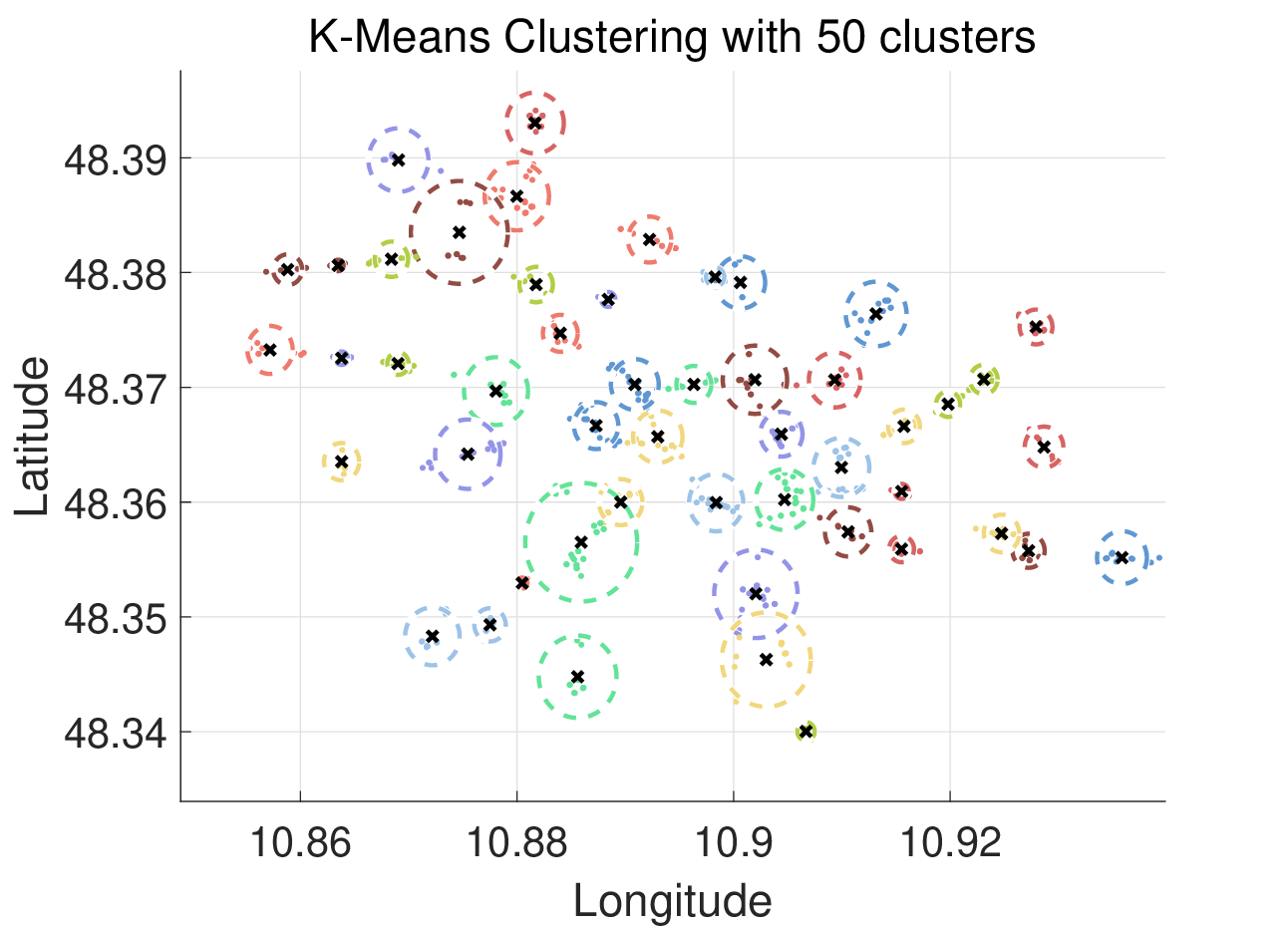}
		\label{fig:k-means50}
	}
	\vspace{-0.4cm}
	\caption{Location and cluster of the detectors.}
	\label{fig:k-means30/50}
	\vspace{-0.3cm}
\end{figure}
We begin by focusing on the real-world UTD19 dataset \cite{Understanding traffic capacity of urban networks}, which contains at least two of the three fundamental traffic variables: speed, flow, and density, for each city. 
Density is typically expressed in terms of detector‐occupancy levels, i.e., the fraction of time each sensor registers a vehicle within the observation window. We employ 
high-resolution traffic data 
from the city of Augsburg 
(Fig. \ref{fig:detmap}), while considering two deployment scenarios with different ES density: one comprising 30 ESs with coverage radii uniformly distributed between 300 m and 700 m (Fig. 
\ref{fig:k-means30}), and the other comprising 50 ESs with radii ranging from 50 m to 400 m (Fig. \ref{fig:k-means50}). After delineating detection points according to lane markings and local traffic volume, we aggregated detection points into coverage zones via k‑means clustering, which serve as the coverage areas of RSUs, based on their geographic coordinates and inter‑sensor distances \cite{K-means}. 
The resulting clusters, 
which define the effective sensing regions, are illustrated in Figs. \ref{fig:k-means30} and \ref{fig:k-means50}.

\vspace{-0.25cm}
\subsubsection{Prediction Performance}
\begin{figure}[]
	\vspace{-5.4mm}
	\centering
	\subfigure[Half-hour (25th)]{
	\includegraphics[ width=0.3\columnwidth]{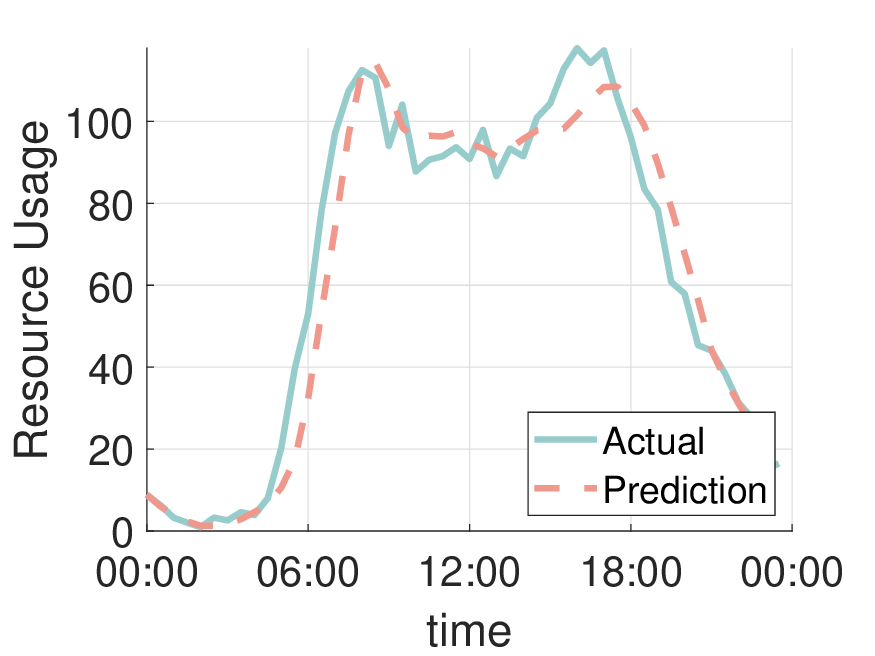} 
	\label{fig:LSTM502}
	}	
	\subfigure[Half-hour (30th)]{
	\includegraphics[ width=0.3\columnwidth]{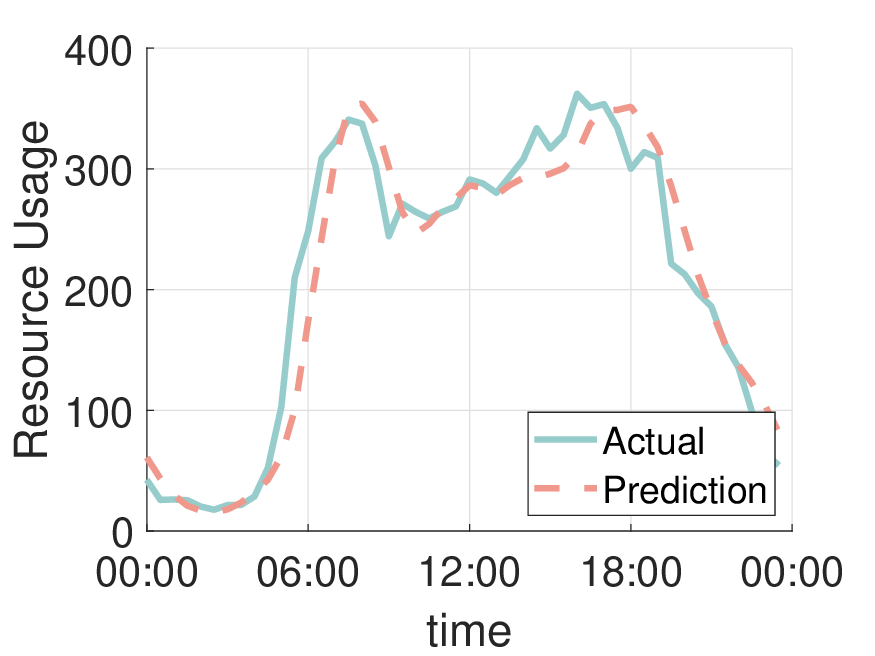} 
	\label{fig:LSTM501}
	}
	\subfigure[Half-hour (35th)]{
	\includegraphics[ width=0.3\columnwidth]{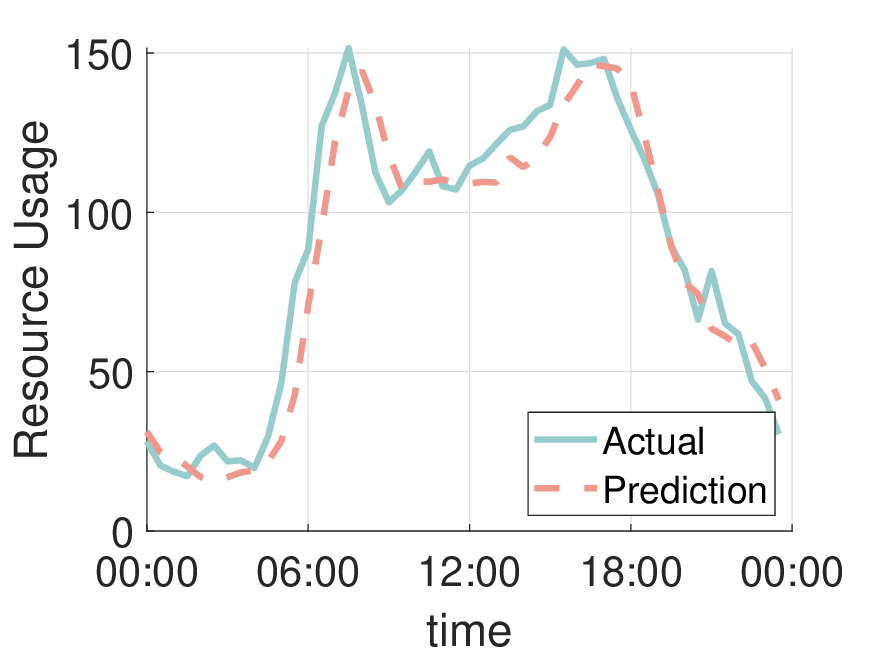} 
	\label{fig:LSTM50}
	}
	\subfigure[One-hour (25th)]{
	\includegraphics[ width=0.3\columnwidth]{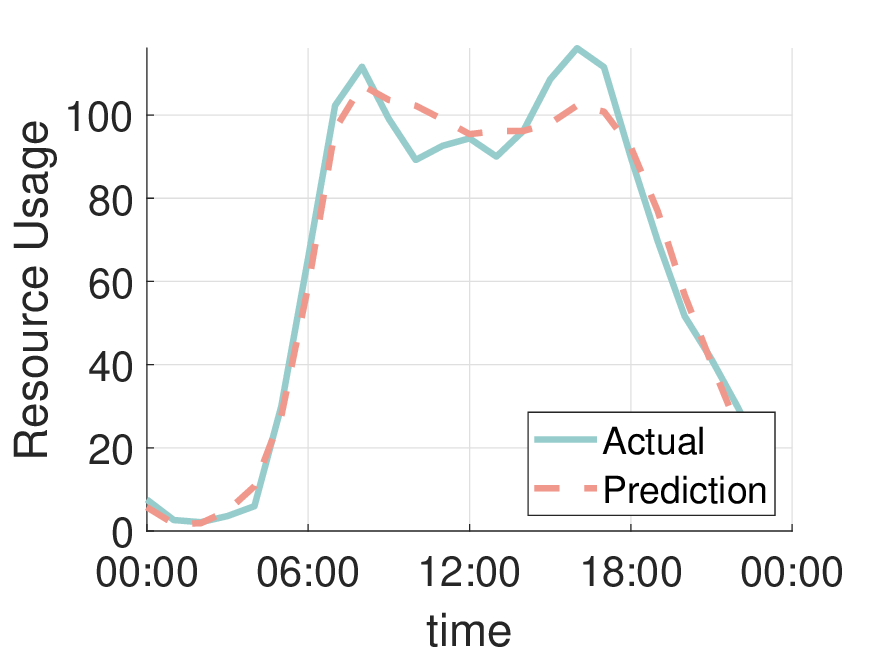}    
	\label{fig:LSTM302}
	}
		\subfigure[One-hour (30th)]{
		\includegraphics[ width=0.3\columnwidth]{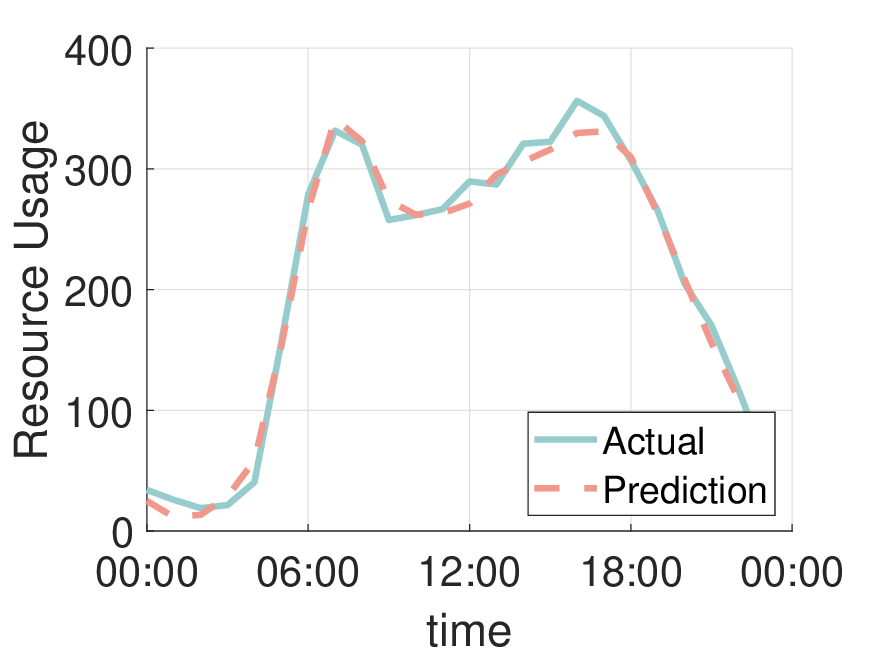}    
		\label{fig:LSTM301}
	}
	\subfigure[One-hour (35th)]{
	\includegraphics[ width=0.3\columnwidth]{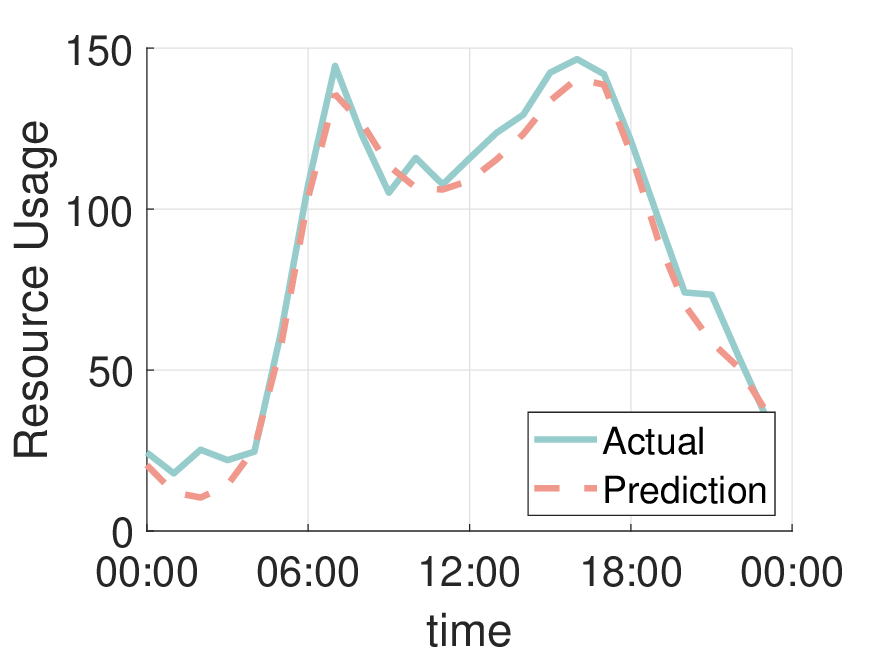}    
	\label{fig:LSTM30}
	}
    	\vspace{-3.6mm}
	\caption{Predicted resource usage vs. actual data.}
	\label{fig:LSTM30_50}
	\vspace{-0.7cm}
\end{figure}

We construct historical resource‑usage profiles for each cluster using traffic data collected from May 9 to May 16, 2017. These profiles incorporate key characteristics of each coverage zone, namely, its geographical area, number of lanes, and prevailing vehicle speeds. Leveraging these week‑long profiles, we forecast the next day’s resource demand for each cluster (i.e., each ES’s coverage area). For example, Fig. \ref{fig:LSTM30_50} depicts the predicted versus actual resource usage for the clusters with index 25, 30 and 35 (out of 50 ones), evaluated at two different temporal granularities.

We do not report full variance distributions for forecasts; 
 instead, prediction accuracy is illustrated through representative case studies. This is because each ES uses an LSTM-based predictor with locally tuned hyperparameters, preventing a unified variance profile. For modeling simplicity, all forecasts are produced at fixed, uniform intervals, as is customary in time‑series analysis. In practice, however, ESs may adopt different prediction cadences and contract horizons, but for simulation consistency, we standardize all ESs to the same discrete time grid. 
%
\vspace{-0.25cm}
\subsubsection{Evaluation Metrics Performance}
In experiments on real‐world dataset, we feed the LSTM‑based forecasts of resource usage into the LAContracts, while actual traffic measurements drive the execution of resource‐trading transactions. We evaluate two temporal granularities (i.e., hourly and half‑hourly intervals) for both prediction and trading. Moreover, each ES’s hardware configuration (including power consumption, processing capacity, etc.) is tailored to its deployment density and coverage radius, which are summarized in Table \ref{tab:table2}.
\begin{table}[b!]
	\vspace{-6mm}
	\scriptsize
	\caption{	
		Key parameters \cite{simulationcover,simulationenergy}\label{tab:table2}}
	\vspace{-2.65mm}
	\centering
	\begin{tabular}{|>{\centering\arraybackslash}m{2.5cm}|@{\hskip 3pt}|c|@{\hskip 3pt}|c|}
		\hline
		\rowcolor{verylightgray}
		\bf{Parameter} & \bf{30 ESs} & \bf{50 ESs}\\
		\hline
		\hline
		Covering Radius & [100,700]m  &  [50,400]m\\
		\hline
		\rowcolor{verylightgray}
		$R^\text{In}$ & [200,500] & [50,300] \\
		
		\hline
		$\eta^\text{Use}$ & [200,500]W/RB &[100,300]W/RB\\
		
		\hline
		\rowcolor{verylightgray}
		$\eta^\text{Idle}$ & [20,30]W & [10,20]W \\
		\hline	
		$\omega_{i,n}^\text{B}$ & [0.2-0.3]W/bite/m & [0.1-0.2]W/bite/m \\
		
		%
		
		\hline
	\end{tabular}
    \vspace{-3mm}
\end{table}
\begin{figure}[t]
\vspace{-5.4mm}
	\centering
	\subfigure[]{
		\includegraphics[trim=0.2cm 0cm 1.5cm 0cm, clip, width=0.3\columnwidth]{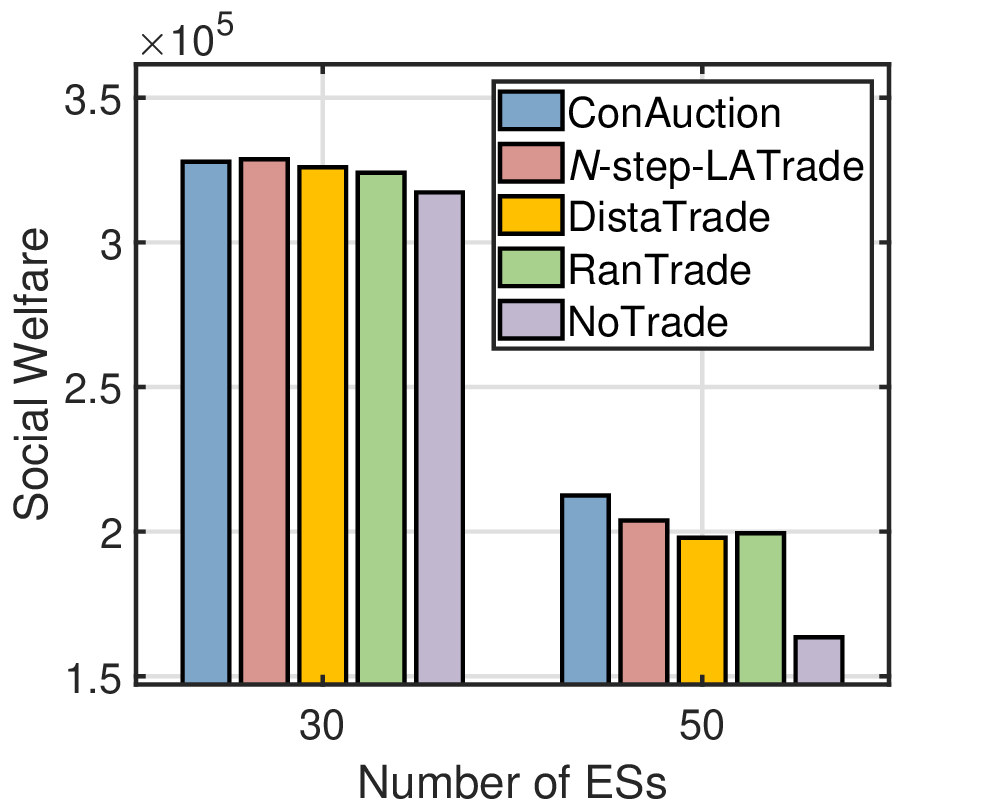}   
		\label{fig:one_SW}     
	}
	\subfigure[]{
		\includegraphics[trim=0.2cm 0cm 1.5cm 0cm, clip, width=0.3\columnwidth]{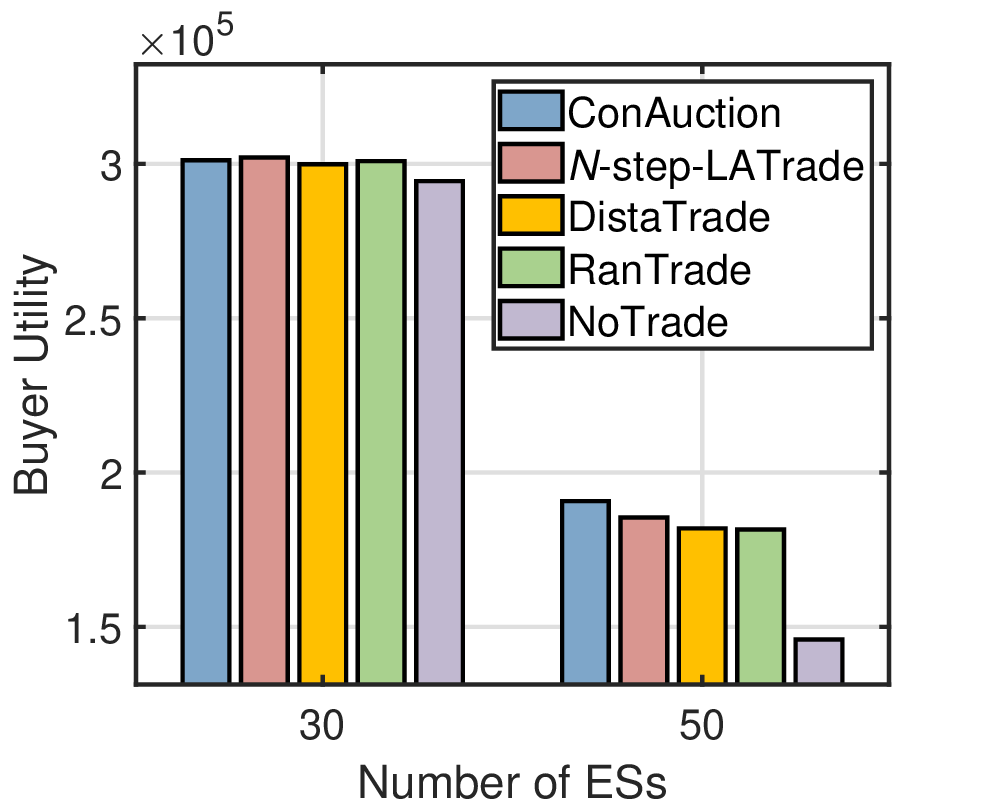}  
		\label{fig:one_BU}      
	}
	\subfigure[]{
		\includegraphics[trim=0.2cm 0cm 1.5cm 0cm, clip, width=0.3\columnwidth]{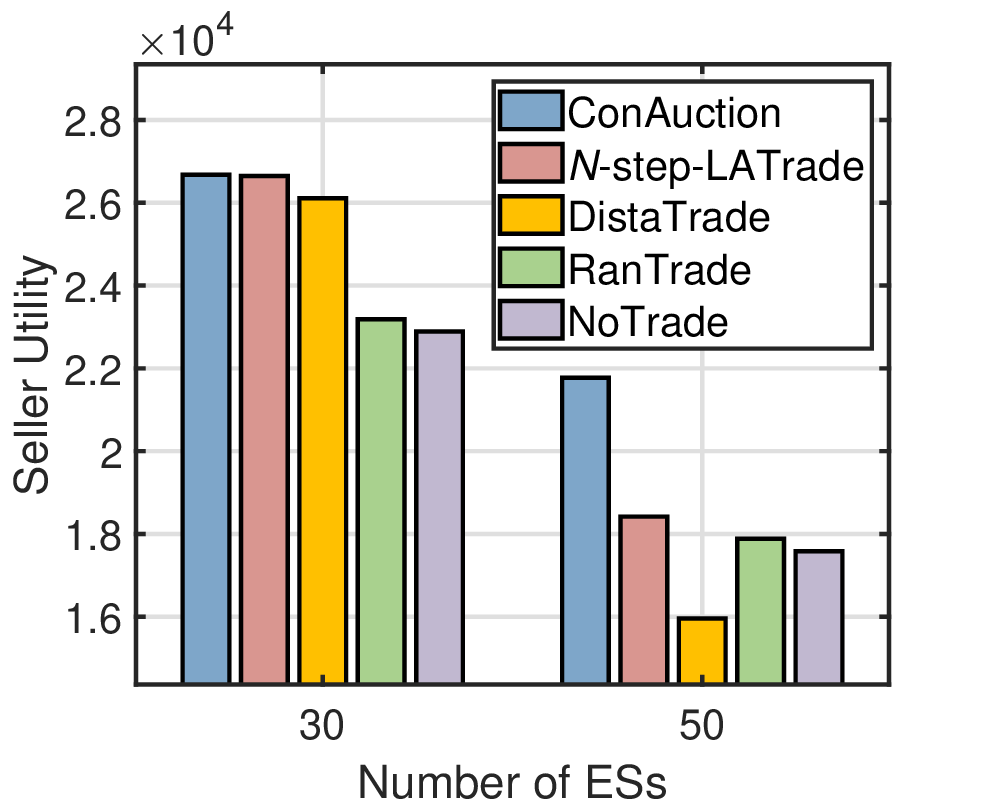}
		\label{fig:one_SU}       
	}
	\\[-3mm]
	\subfigure[]{
		\includegraphics[ trim=0.2cm 0cm 1.5cm 0cm, clip,width=0.3\columnwidth]{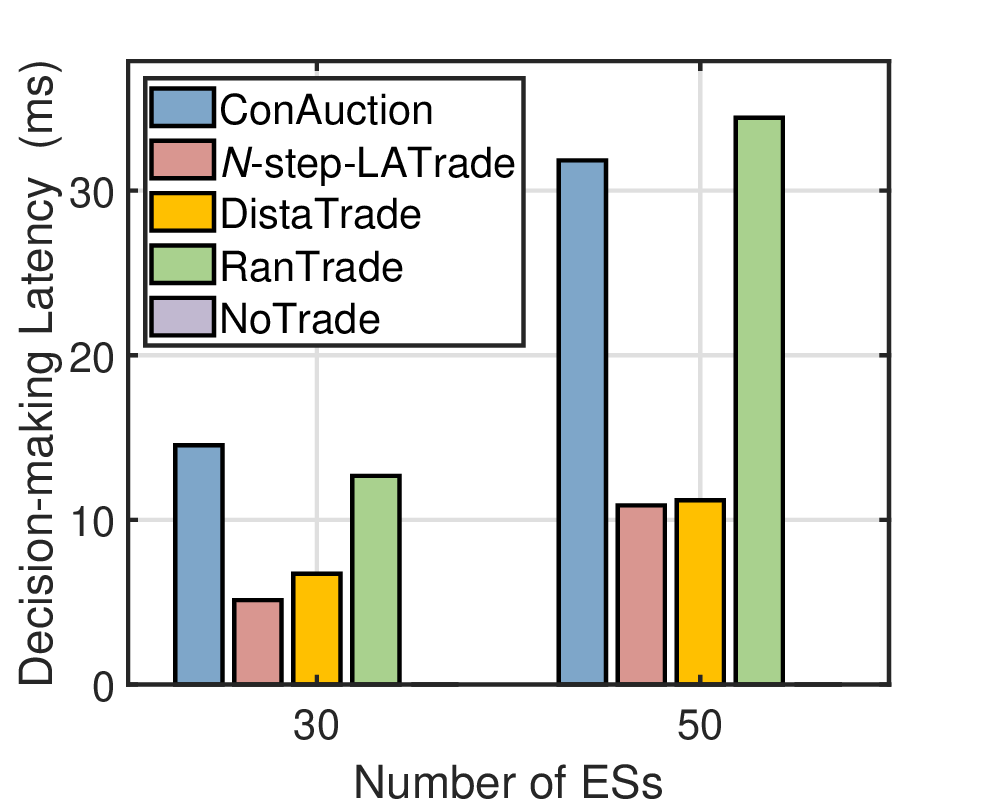}
		\label{fig:one_ET} 
	}
	\subfigure[]{
		\includegraphics[trim=0.2cm 0cm 1.5cm 0cm, clip, width=0.3\columnwidth]{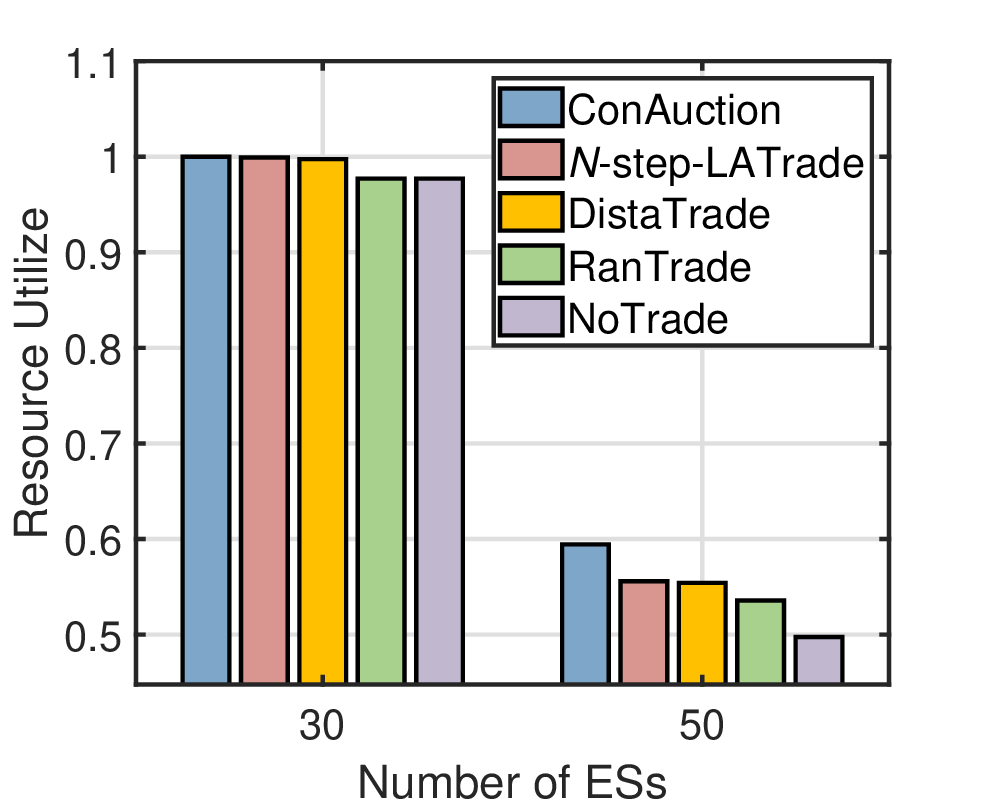}
		\label{fig:one_RU} 
	}
	\subfigure[]{
		\includegraphics[ trim=0.2cm 0cm 1.5cm 0cm, clip,width=0.3\columnwidth]{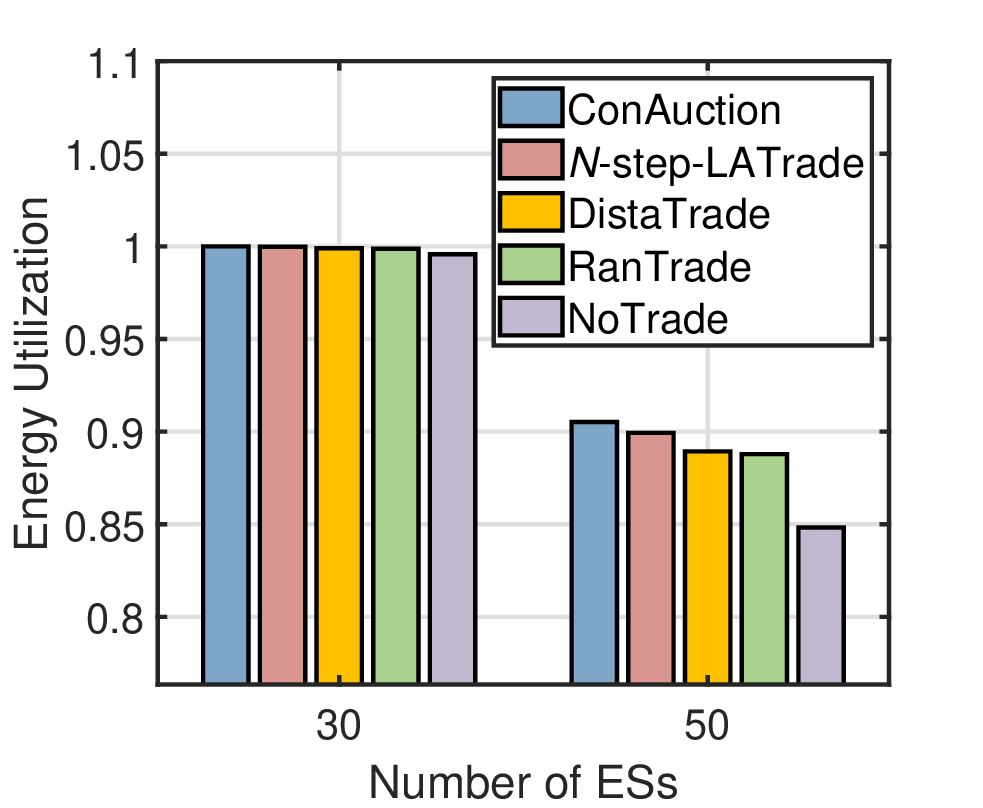}
		\label{fig:one_EU} 
	}
	\vspace{-4.5mm}
	\caption{Performance measured on the real-world dataset with  one-hour-interval prediction.}
	\label{fig:realWorlddataOnehour}
	\vspace{-0.65cm}
\end{figure}
In Fig. \ref{fig:realWorlddataOnehour}, Figs. \ref{fig:one_SW}-\ref{fig:one_SU} present the comparison of social welfare and individual utilities of buyers and sellers under two ES deployment settings: 30 and 50 ESs.
In Fig. \ref{fig:one_SW}, under the 30-ES setting, our proposed method, $N$-step-LATrade, achieves social welfare that is very close to ConAuction, and outperforms other baselines including DistaTrade, RanTrade, and NoTrade. This result demonstrates that our mechanism enables ESs, acting as buyers, to secure cheaper services in advance, potentially at lower costs than executing tasks locally. As a result, the execution cost is reduced, which directly improves the total social welfare. In contrast, methods such as DistaTrade (prioritizes nearby partners without considering pricing) and RanTrade (random matching) fail to fully exploit cost-saving opportunities.
Under the 50 ES deployment, our method performs slightly worse than ConAuction, but still better DistaTrade, RanTrade, and NoTrade. This small gap is expected, as our contracts are pre-signed based on predicted supply and demand, while ConAuction benefits from real-time information during on-site trading.
It is also evident that the overall social welfare in the 30 ES setting is higher than in the 50 ES setting across all methods. This is because the total city coverage is fixed, and thus as the number of ESs increases, each ES covers a smaller area, resulting in shorter transmission ranges and more overlap in their peak and off-peak periods. These effects reduce the likelihood of beneficial trades, thus lowering the total transaction volume and welfare.
Figs. \ref{fig:one_BU} and \ref{fig:one_SU} show buyer and seller utilities, respectively. While these metrics alone are not our primary focus, they confirm that $N$-step-LATrade significantly outperforms NoTrade in both cases, demonstrating its ability to incentivize ESs to participate in the resource market and benefit from collaboration.
Similar results are also observed in Figs. \ref{fig:half_SW}--\ref{fig:half_SU}.

Since time efficiency is one of the most important considerations in highly dynamic networks such as IoV, in Fig. \ref{fig:one_ET} we depict the delay of trading decision-making upon considering different numbers of ESs.  
As shown, the latency introduced by resource trading increases as the number of ESs increases. Also, in the absence of trading (i.e., NoTrade), tasks are dispatched immediately, incurring no additional delay. However, when ConAuction is employed, each ES must negotiate and match with a partner, in real-time, resulting in substantial overhead. For example, in Fig. \ref{fig:one_ET}, the decision-making latency reaches approximately 15 ms for 30 ESs, and rises to 30 ms for 50 ESs, which can be prohibitive for delay-sensitive applications\footnote{In time-critical IoV applications such as autonomous driving or cooperative collision avoidance, end-to-end latency requirements are extremely stringent, to ensure real-time responsiveness and safety \cite{latency1}.}. In contrast, our proposed $N$-step-LATrade reduces this latency to roughly 5 ms for 30 ESs and only marginally higher for 50, thus demonstrating a threefold improvement over ConAuction and validating the time efficiency gains of pre-negotiated agreements. The same trends can be observed in Fig. \ref{fig:half_ET}.


Finally, Figs. \ref{fig:one_RU} and \ref{fig:one_EU} depict resource and energy utilization under both deployment densities. Regarding 30 ESs scenario, the resource utilization of our method reaches its maximum, as driven by a high trading-success rate, while with 50 ESs it remains above 55\%, representing a 10\% improvement over DistaTrade, RanTrade, and NoTrade. Energy utilization mirrors this trend: because active operation draws significantly more power than idle standby, the energy-use rate exceeds the resource-use rate. Importantly, our pre-auction-enabled framework consistently surpasses the no-trading case, markedly reducing both resource and energy waste. Figs. \ref{fig:half_RU} and \ref{fig:half_EU} further corroborate these results across hourly and half-hourly time resolutions, underscoring the robustness of our forecasting and trading mechanisms.

\vspace{-0.4cm}
\subsection{Synthetic Data-driven Experiments}
To evaluate scalability and generalizability of our method, we next perform numerical simulations on a synthetic dataset. 
We vary the number of ESs from 10 to 50, placing them uniformly at random within an area that grows from 1 to 6.5 km² as the number of ESs increases. The ESs’ coverage radii are drawn from 200 m to 700 m.
Each ES is assigned an inherent resource-block capacity, $R^\text{In}$, sampled uniformly at random as an integer between 50 and 200 to reflect its supply. We then model the additional resource demand by fitting a truncated normal distribution whose mean equals $R^\text{In}$ and whose variance is chosen uniformly between 10 and 60. From this truncated density, we compute the expected additional resource blocks, $R^\text{Est-}$, and hence the total predicted capacity, ${R}^\text{Est}$. Finally, the actual resource usage, $R^\text{Act}$, is drawn randomly from the same truncated distribution, allowing for deviations between predicted and realized demand.
The simulated market further incorporates transaction economics by assigning each ES a per‑RB revenue drawn uniformly from 50 to 100 monetary units, while remaining idle incurs a loss sampled uniformly between 5 and 15 units. Other parameter values are set according to Table 1.
%
\begin{figure}[b]
	\vspace{-5.4mm}
	\centering
	\subfigure[]{
		\includegraphics[trim=0.2cm 0cm 1.5cm 0cm, clip, width=0.3\columnwidth]{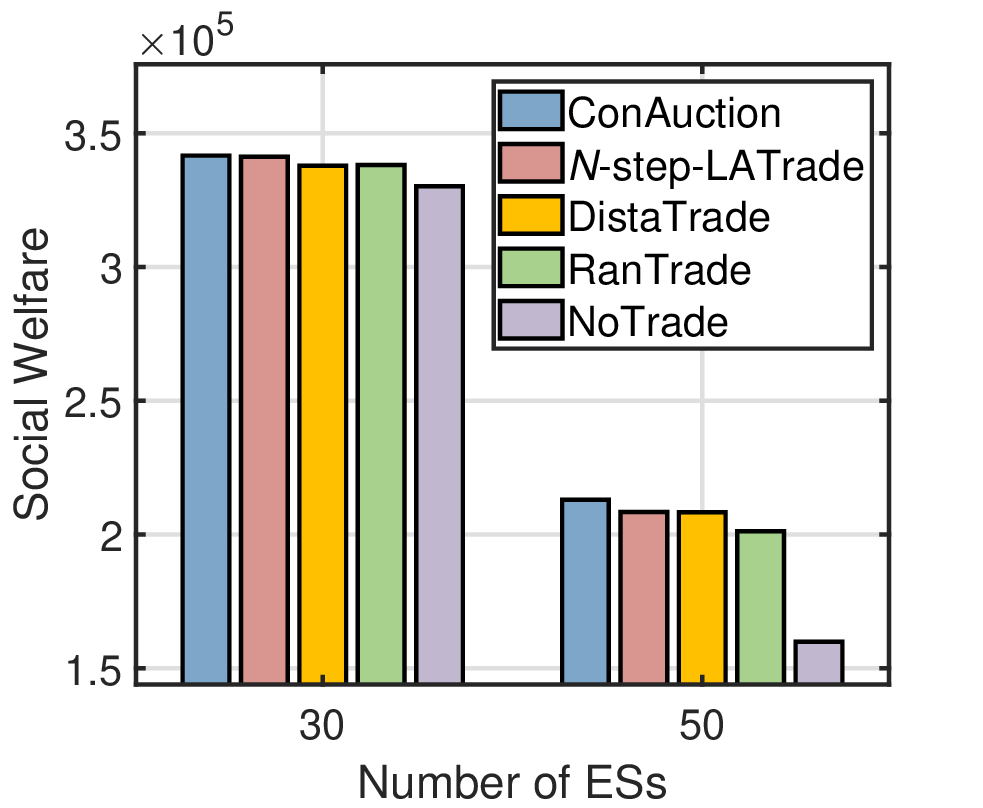}      \label{fig:half_SW}
	}
	\subfigure[]{
		\includegraphics[ trim=0.2cm 0cm 1.5cm 0cm, clip, width=0.3\columnwidth]{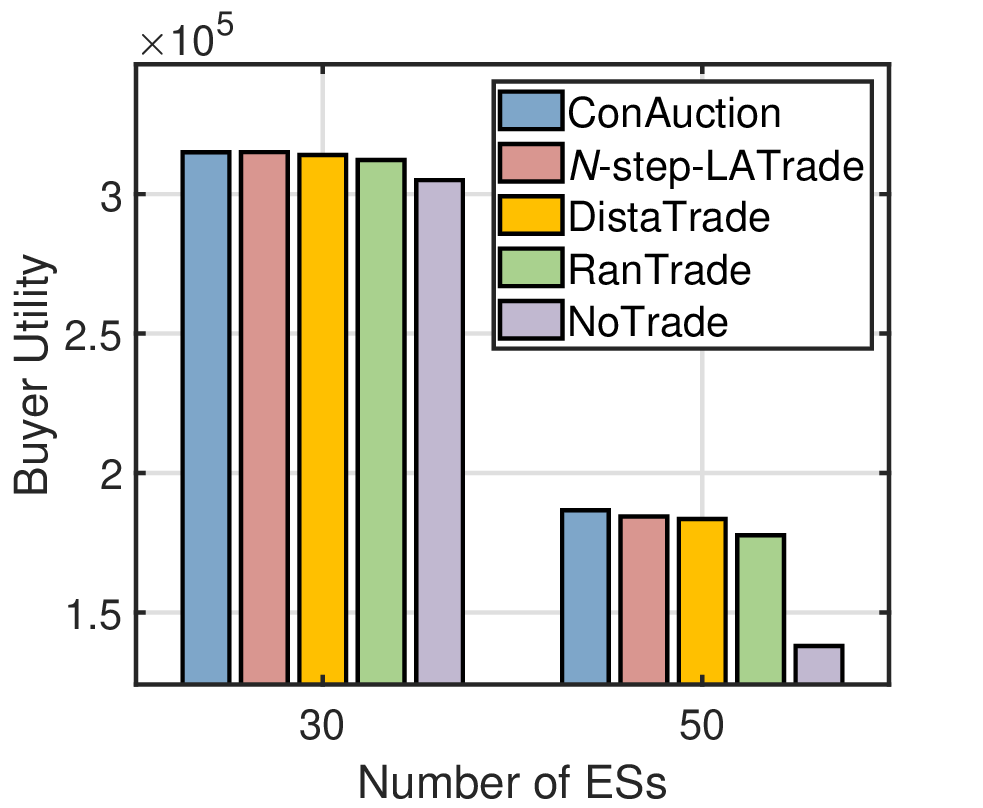}
		\label{fig:half_BU}       
	}
	\subfigure[]{
		\includegraphics[trim=0.2cm 0cm 1.5cm 0cm, clip, width=0.3\columnwidth]{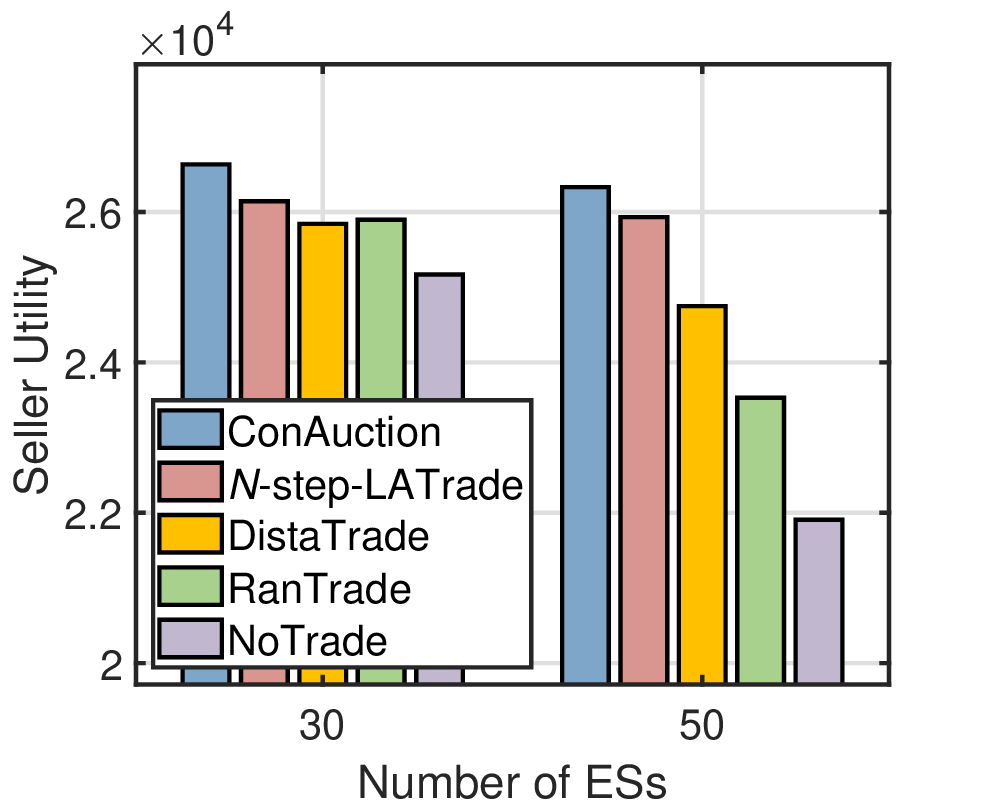}      \label{fig:half_SU}
	}
	\\[-3mm]
		\subfigure[]{
			\includegraphics[ trim=0.2cm 0cm 1.5cm 0cm, clip,width=0.3\columnwidth]{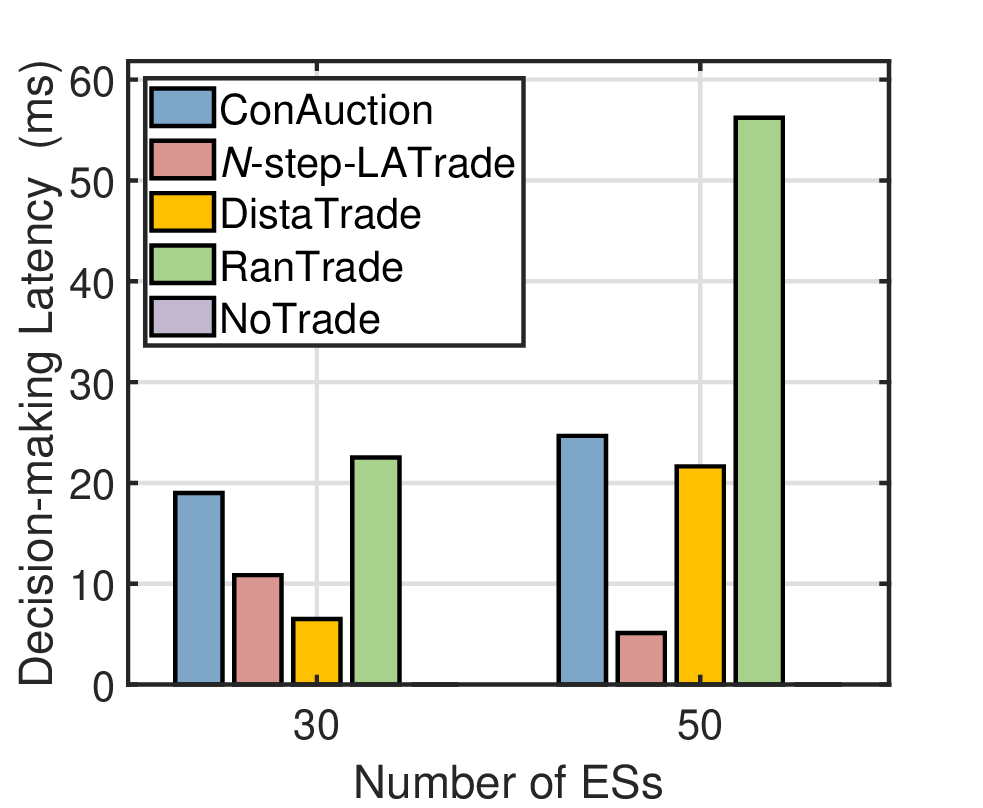}
			\label{fig:half_ET}
		}
			\subfigure[]{
				\includegraphics[trim=0.2cm 0cm 1.5cm 0cm, clip, width=0.3\columnwidth]{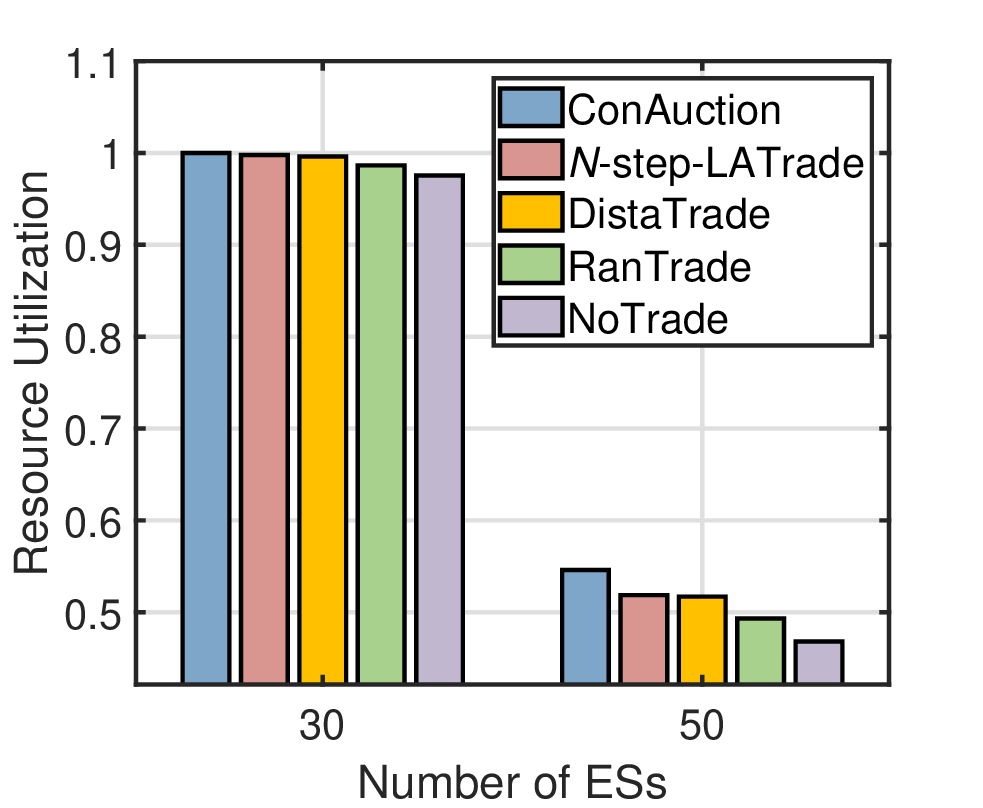}
				\label{fig:half_RU}
			}
				\subfigure[]{
					\includegraphics[ trim=0.2cm 0cm 1.5cm 0cm, clip,width=0.3\columnwidth]{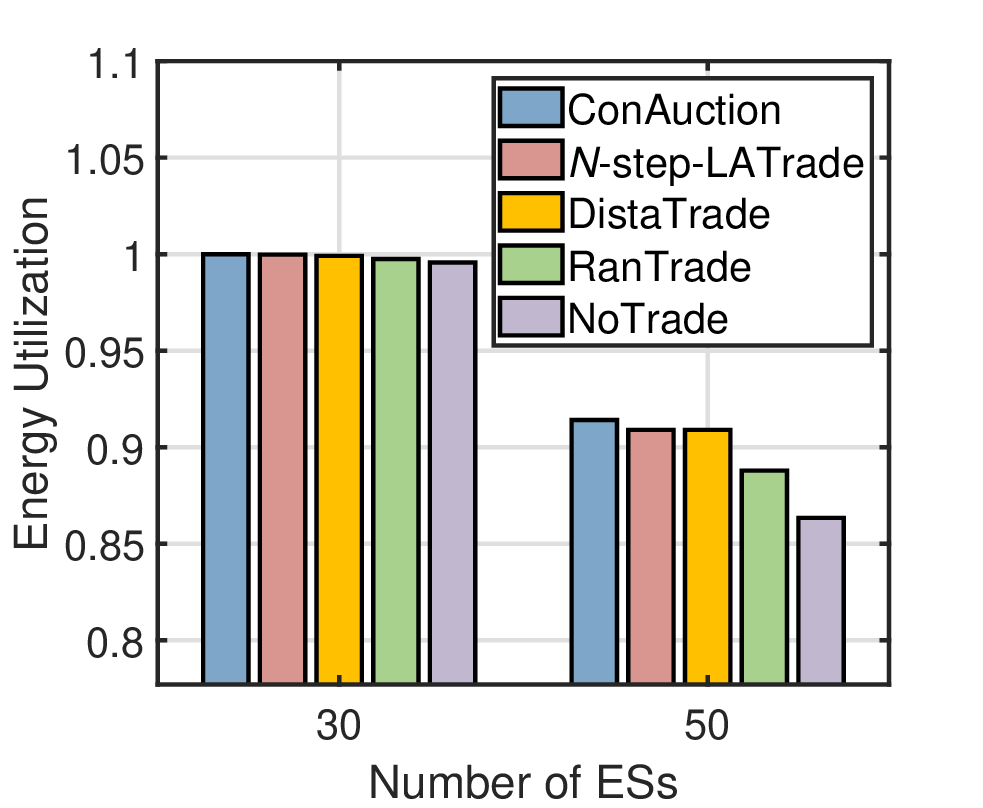}
					\label{fig:half_EU}
				}
				\vspace{-4.5mm}
				\caption{Performance measured on the real-world dataset with  half-hour-interval prediction.}
				\label{fig:realWorlddataHalfhour}
				\vspace{-0.65cm}
			\end{figure}
\begin{figure}[]
	\centering
%
		\centering
		\subfigure[]{
			\includegraphics[ trim=0.2cm 0cm 1.2cm 0cm, clip,width=0.3\columnwidth]{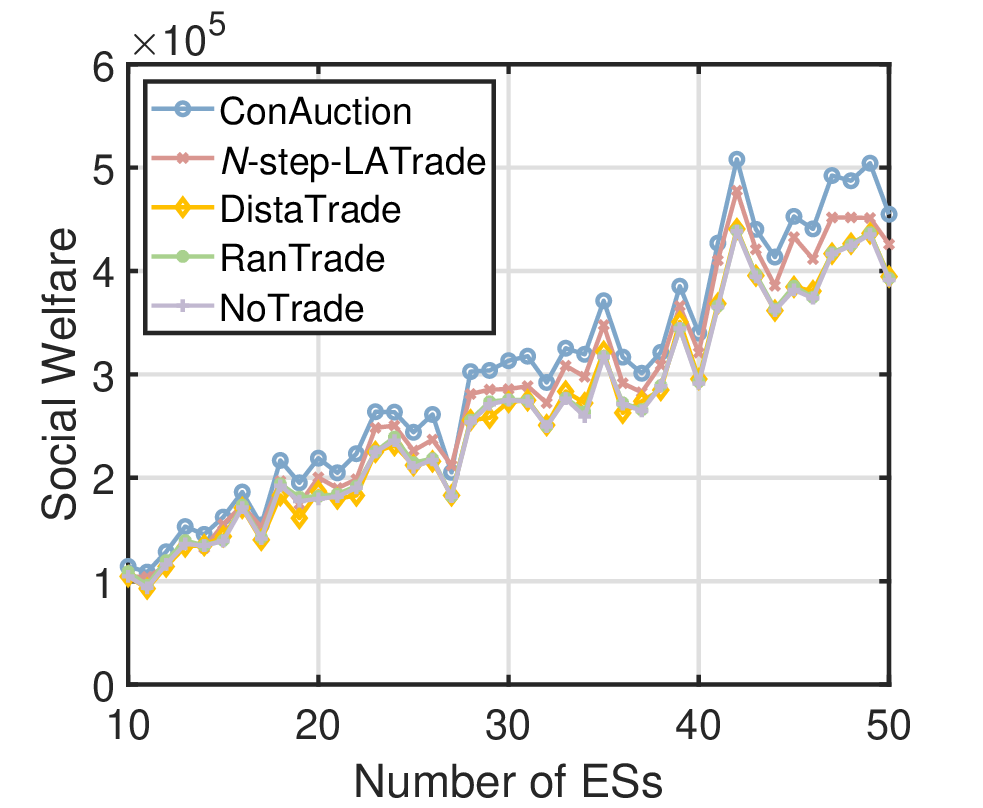} 
			\label{fig:numerical_a}
		}
		\subfigure[]{
			\includegraphics[ trim=0.2cm 0cm 1.2cm 0cm, clip,width=0.3\columnwidth]{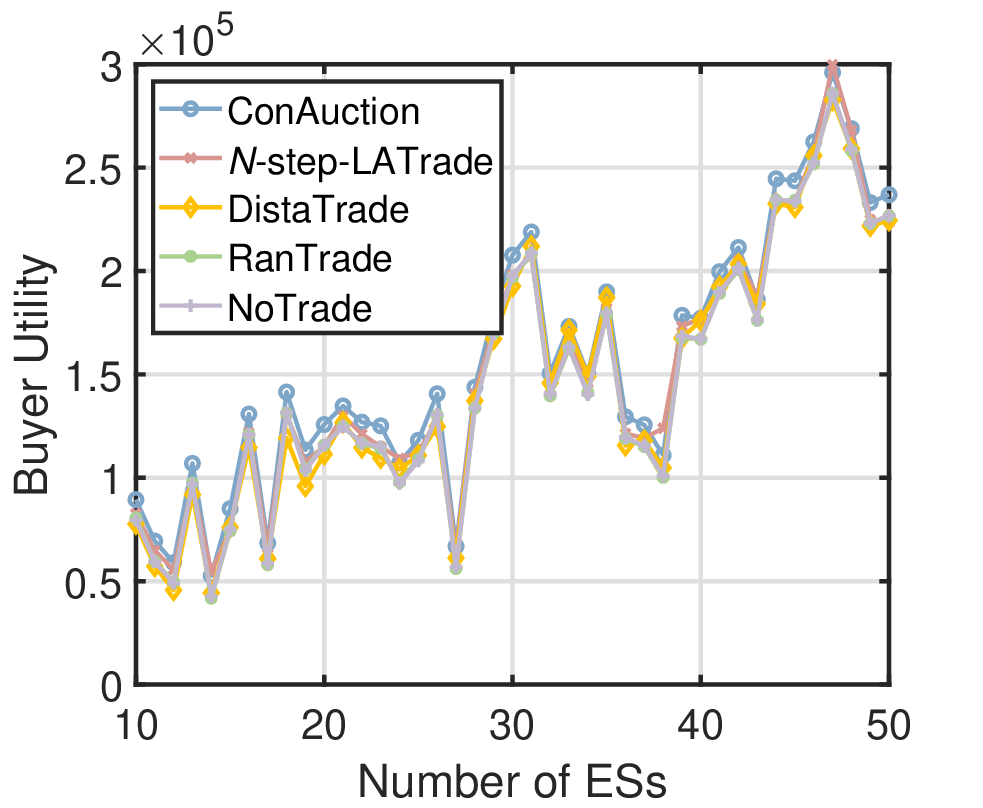} 
			\label{fig:numerical_1}
		}
		\subfigure[]{
			\includegraphics[ trim=0.2cm 0cm 1.2cm 0cm, clip,width=0.3\columnwidth]{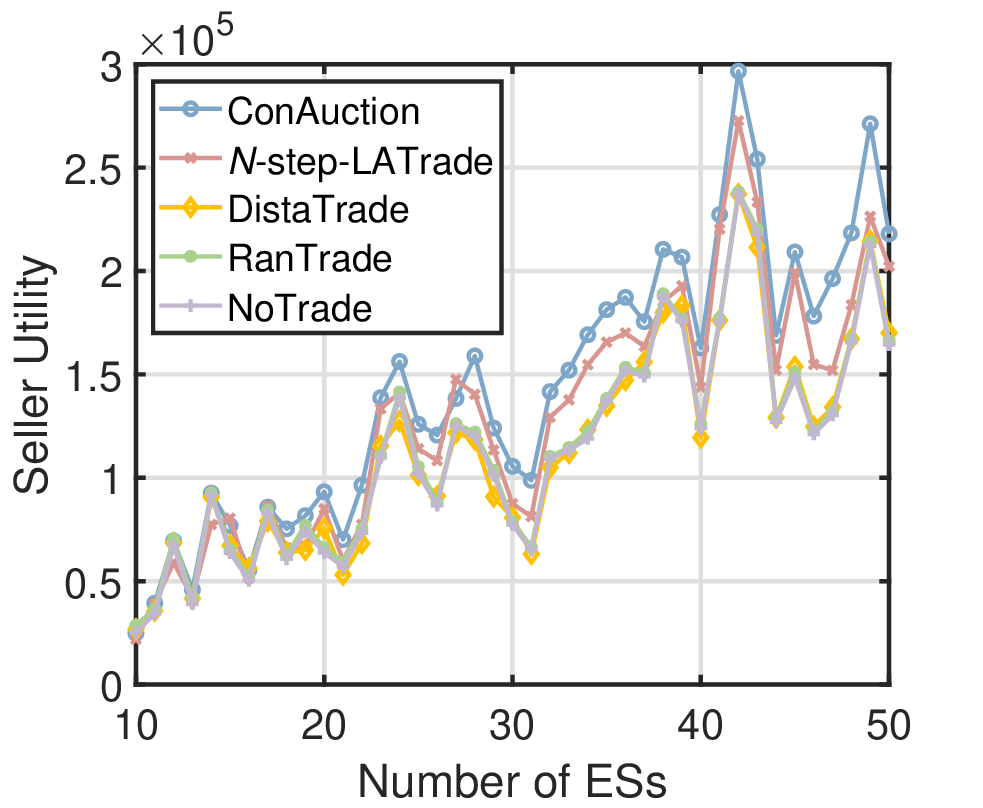} 
			\label{fig:numerical_2}
		}
		\\[-3mm]
		\subfigure[]{
			\includegraphics[ trim=0.2cm 0cm 1.2cm 0cm, clip,width=0.3\columnwidth]{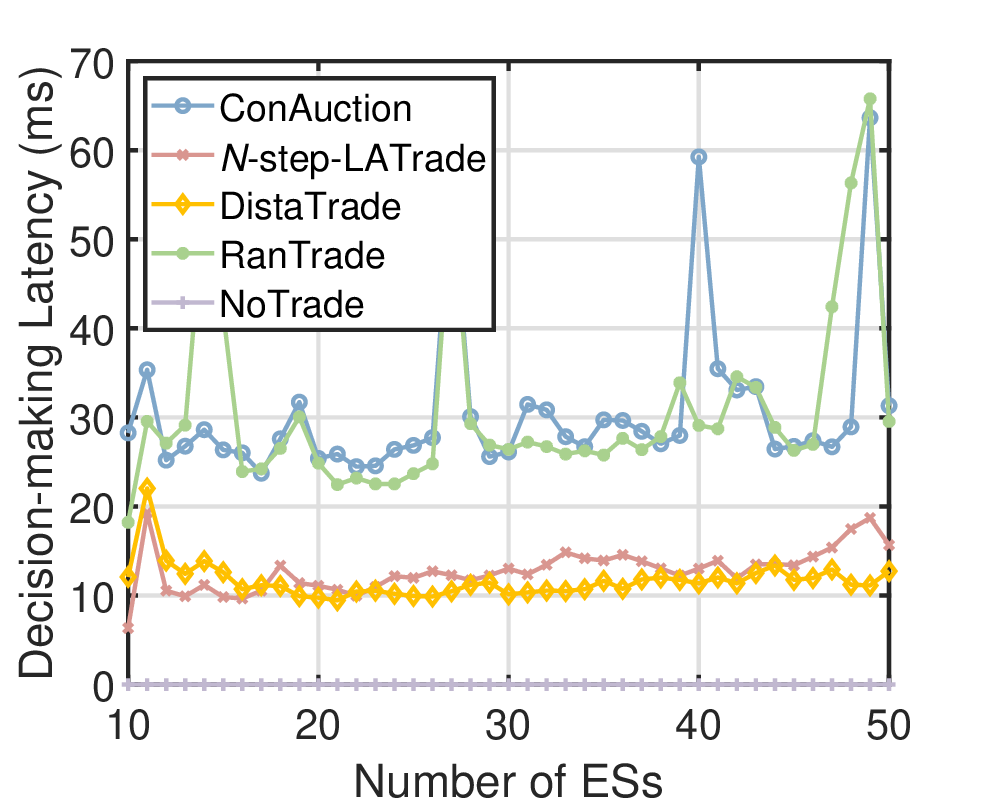}
			\label{fig:numerical_b}
		}
		\subfigure[]{
			\includegraphics[ trim=0.2cm 0cm 1.2cm 0cm, clip,width=0.3\columnwidth]{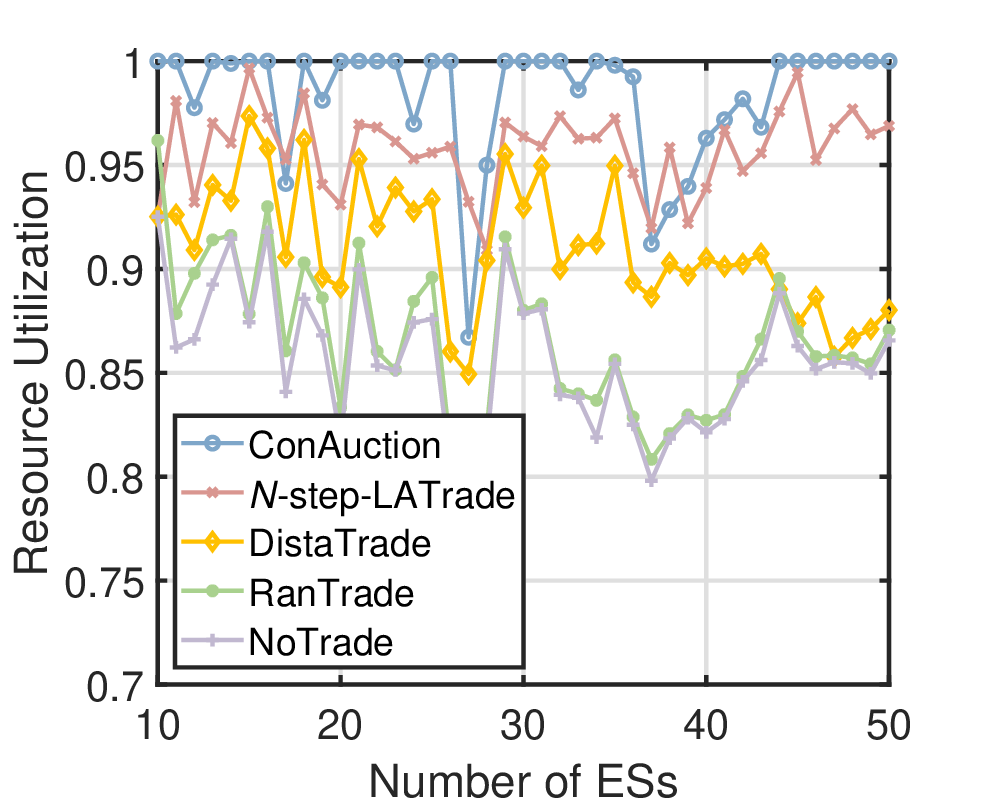}
			\label{fig:numerical_c}
		}
		\subfigure[]{
			\includegraphics[ trim=0.2cm 0cm 1.2cm 0cm, clip,width=0.3\columnwidth]{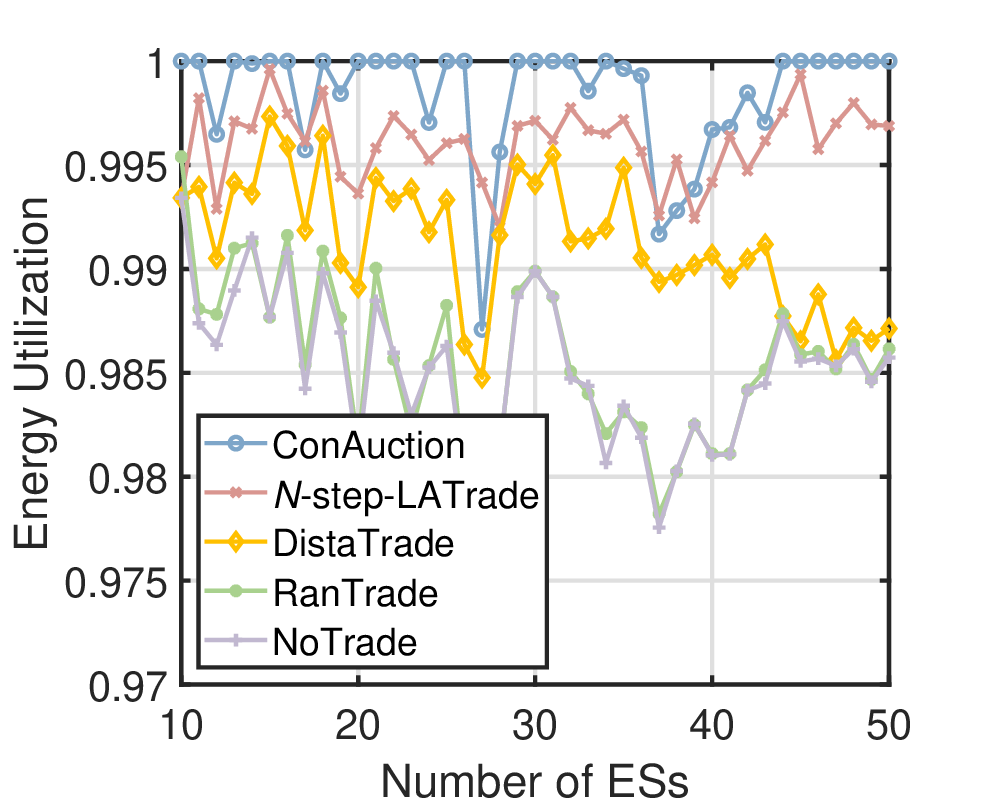}
			\label{fig:numerical_d}
		}
	\vspace{-0.3cm}
	\caption{Performance evaluation on synthetic data-driven experiments.}
	\label{fig:numerical}
	\vspace{-0.7cm}
\end{figure}

Figs. \ref{fig:numerical_a}-\ref{fig:numerical_2} summarize the social welfare and utility outcomes across all simulated scenarios. As expected, the NoTrade method consistently yields the poorest social welfare, while ConAuction achieves the highest. Our proposed $N$-step-LATrade ranks a close second, outperforming DistaTrade method, which minimizes transmission cost but ignores bid-ask matching and therefore delivers suboptimal welfare. Although random on-site trading can, in rare high-volatility cases, temporarily surpass contract-based welfare, its average performance remains below that of our $N$-step-LATrade. Utility distributions mirror these trends at the individual-ES level, demonstrating that our contract formulation not only maximizes aggregate welfare but also delivers more equitable and predictable gains to each participant.

Fig. \ref{fig:numerical_b} presents the time efficiency across different methods. As a reminder, the NoTrade method incurs zero latency since it involves no inter-ES communication or data exchanges. Among other approaches, ConAuction exhibits the highest delay due to its on-site matching and bidding processes, whereas our $N$-step-LATrade approach achieves the second-lowest latency, limited to the overhead of transaction execution and default reporting. RanTrade method, despite omitting formal matching and price determination phases, still suffers from 30-50 ms of delay from partner selection and pricing checks, which can be unacceptable for ultra-latency-sensitive tasks. In contrast, our $N$-step-LATrade consistently finishes its execution in approximately 10 ms.

Finally, Figs. \ref{fig:numerical_c}-\ref{fig:numerical_d} depict resource and energy utilization, which are defined respectively as the fraction of computational capacity and active power draw relative to their totals, across varying ES counts. As expected, ConAuction achieves the highest resource utilization, followed by our $N$-step-LATrade and the DistaTrade, both of which implement advance transactions. RanTrade yields only marginal improvements over the no‑trading case. Resource utilization improves with the number of successful buyer-seller pairings in each interval, directly reducing idle capacity. Energy utilization tracks this trend: fewer idle resources translate to lower standby energy losses, so higher resource use corresponds to more of the energy budget being devoted to productive processing. Altogether, these results confirm that our $N$-step-LATrade not only sustains near‑optimal resource utilization but also curtails energy waste.

\vspace{-4mm}
\section{Conclusion and Future Work}
\vspace{-.4mm}
This paper tackled dynamic resource exchanges in EdgeIoV environments by introducing a stagewise, prediction-assisted contract auction framework. Through spatio-temporal demand forecasting and the construction of $N$-step look-ahead service contracts, our approach enables ESs to proactively determine their trading roles and establish future transactions with predefined terms. 
Evaluations on both real-world and synthetic datasets  demonstrated the effectiveness of our method.
Future work can aim to migrate from abstract computing tasks to supporting multi-modal foundation model execution at the vehicles, transforming auctions into enablers of large-scale, intelligent services. This entails designing \textit{contract markets for collaborative machine learning (ML)}, where ESs not only trade computing resource but also exchange model updates to train heavy foundation models.

\newpage
\clearpage
\appendix
\section*{Property Analysis}\label{proof}
Our work offers a novel approach by incorporating future resource usage predictions and an advance double auction mechanism that looks $N$ steps ahead for resource trading. Consequently, these properties should hold not only at the contract signing phase (i.e., at $t_0$) but also remain valid throughout the entire time horizon, encompassing all transactions from a long-term perspective.

\theoremstyle{plain}
\setcounter{theorem}{0} 
\begin{theorem}
	The proposed methodology ensures individual rationality for all the ESs.
\end{theorem}
As previously discussed, we divide the entire time domain into multiple time frames, which are further grouped into two stages: one corresponding to the initial stage at $t_0$, and the other comprising the remaining time frames. As outlined in Definition \ref{def:definition1}, we demonstrate individual rationality from three perspectives, detailed in Lemmas \ref{lemma:1.1} to \ref{lemma:1.3}. The first perspective pertains to the pre-auction stage at $t_0$, where we show that ESs act rationally when participating in the double auction process. The remaining two perspectives focus on the execution phase (i.e., from 
$t_1$ to $t_n$), demonstrating that ESs remain individually rational while implementing the pre-signed LAContracts during practical transactions.

\begin{lemma}\label{lemma:1.1}
	The proposed double auction during $t_0$ holds individual rationality for all ESs acting as either buyers or sellers.
\end{lemma}
\begin{proof}
	To show that our designed pre-double auction process for LAContracts signing is individual rational, we take the transaction during $t_n$ as an example. 
	
	\noindent
	$\bullet$ \textit{For sellers,} if $\bm{s}_{j,n}$ has not been mapped to any buyer, we definitely have  $p_{i,j,n} = 0$. Accordingly, there is no contract signed from the seller’s side, and thus no trading between this seller and any buyers during the practical transaction. In other words, the seller will neither earn income nor incur any loss from interactions with buyer ESs.  Then, if $\bm{s}_{j,n}$ has been matched to a buyer, the trading price determined in Alg. 2 can meet $\mathsf{ask}_{j,n}\leq p_{i,j,n}$ and $\mathsf{ask}_{j,n}\leq p^\text{S}_{j,n}$. Similarly, for buyers, if a buyer loses in the designed auction, it makes no payment and is not bound by any contract. When a buyer is successfully matched with a seller, the trading price specified in the LAContract satisfies $p_{i,j,n}\leq \mathsf{bid}_{i,j,n}$ and $p^\text{B}_{i,n}\leq \mathsf{bid}'_{i,n}$, ensuring that the buyer’s total payment, including data transmission costs, remains below or equal to its submitted bids. Therefore, the proposed methodology upholds the principle of individual rationality for ESs during the initial contract-signing stage at $t_0$.
\end{proof}
\begin{lemma}\label{thm:theorem2}
	Implementing the pre-signed LAContracts during practical transactions (i.e., from $t_1$ to $t_N$ ) guarantees non-negative profits for ESs.
\end{lemma}
\begin{proof}
	Given our unique approach of anticipating future transactions by encouraging ESs to sign LAContracts at time $t_0$, the execution of these contracts in later stages inevitably encounters uncertainties due to the dynamic and stochastic nature of the market. For instance, inaccurate predictions of future resource usage may lead to contract breaches, resulting in penalties and potential losses for buyers. Therefore, we proceed to analyze the property of individual rationality for both buyers and sellers across all practical transactions, i.e., for every $t_n$ where $n>0$. A critical scenario arises when a buyer's realized utility becomes negative due to having to paying compensation for defaulting on a contract, as expressed in equation \eqref{risk}.
	
	\noindent
	$\bullet$ \textit{For a buyer,} regarding $t_n$ as an example, although the buyer may still receive income from its own covered vehicles, its net utility can become negative in certain cases—particularly when it breaches a pre-signed LAContract and must pay a penalty to the corresponding seller. As previously discussed in equation \eqref{ExpUbuyers}, such a scenario arises when the condition in equation \eqref{risk} is satisfied.
	\begin{align}\label{risk}
		&(\mathsf{bid}'_{i,n} \times \min(R^{b,\text{Act}}_{i,n}, R^{\text{B},\text{In}}_{i,n}+
		R^{\text{B},\text{Tra}}_{i,n}) 
		\notag\\
		&- (R^{\text{B},\text{Tra}}_{i,n} - \kappa) p_{i,n}^\text{B} -  \kappa q_{i,n}^\text{B}-\lambda \mathcal{E}^\text{B}_{i,n}  ) < 0. 
	\end{align}
	
	Accordingly, to ensure a non-negative utility, it is necessary to make the following risk under control: 
	\begin{align}\label{probility}
		&\Pr\!\!\left(\!\! \kappa \!\! >\! \!
		\frac{\left(\mathsf{bid}'_{i,n} \! \times \!\min \!\left(\!\! R_{i,n}^{\text{B},\text{Act}}, R_{i,n}^{\text{B},\text{In}} \!+\! R_{i,n}^{\text{B},\text{Tra}}\! \right)\! 
			\!- \! R_{i,n}^{\text{B},\text{Tra}} \! p_{i,n}^{\text{B}} \! - \! \lambda \mathcal{E}_{i,n}^\text{B} \! \right)}
		{p_{i,n}^{\text{B}} + q_{i,n}^{\text{B}}} 
		\!\! \right) 
		\notag\\
		&~~~~~~~~~< \xi^\text{B}_{i,n}, 
	\end{align}
	where $ \xi^\text{B}_{i,n} $ denotes a positive threshold. Since it can be inferred from the constraints \eqref{p0_24a} that $R_{i,n}^{\text{B},\text{Tra}} \leq R_{i,n}^{\text{B},\text{Est-}} $ , we rewrite it as \eqref{scale1}, with  $ \kappa = \min \left(R_{i,n}^{\text{B},\text{Tra}} + R_{i,n}^{\text{B},\text{In}} - R_{i,n}^{\text{B},\text{Act}} , 0 \right)$, representing the quantity of default resources (since we consider the case where the buyer's profit decreases, it only occurs when $ \kappa\geq0$). Therefore, we can simplify the maximum and minimum values as \eqref{scale2} accordingly. In the previous calculations, $R_{i,n}^{\text{B},\text{Est-}}$ denotes the number of resource reported  by the buyer. Recall Section  \ref{4.1}, we use ${R}_{i,n}^{\text{B},\text{Est}}$ to denote buyer $\bm{b}_{i,n}$'s estimate resource usage, where  ${R}_{i,n}^{\text{B},\text{Est}} =  R_{i,n}^{\text{B},\text{Est-}} + R_{i,n}^{\text{B},\text{In}}$, as show in \eqref{scale3}. 
	\begin{figure*}[t]
		\begin{equation}\label{scale1}
        \resizebox{0.93\hsize}{!}{$
			\kappa >\frac{ 
				\mathsf{bid}'_{i,n} \times \min \left( R_{i,n}^{\text{B},\text{Act}}, R_{i,n}^{\text{B},\text{In}} + R_{i,n}^{\text{B},\text{Est-}} \right) - R_{i,n}^{\text{B},\text{Est-}} p_{i,n}^{\text{B}}  - \lambda \Delta t_n \left( \min \left( R_{i,n}^{\text{B},\text{In}}, R_{i,n}^{\text{B},\text{Act}} \right) \eta_{i,n}^{\text{B},\text{Use}}  - \max \left( R_{i,n}^{\text{B},\text{In}} - R_{i,n}^{\text{B},\text{Act}}, 0 \right) \eta_{i,n}^{\text{B},\text{Idle}} \right)}
			{p_{i,n}^{\text{B}} + q_{i,n}^{\text{B}} }
            $}
		\end{equation}
        \rule{\linewidth}{0.75pt}
		\begin{align}\label{scale2}
		\hspace{-4mm}	R_{i,n}^{\text{B},\text{Est-}} + R_{i,n}^{\text{B},\text{In}} - R_{i,n}^{\text{B},\text{Act}} >
			\begin{cases}
				\frac{\mathsf{bid}'_{i,n} \times R_{i,n}^{\text{B},\text{Act}} - R_{i,n}^{\text{B},\text{Est-}} p_{i,n}^{\text{B}} - \lambda \Delta t_n \big( R_{i,n}^{\text{B},\text{Act}} \eta_{i,n}^{\text{B},\text{Use}} - (R_{i,n}^{\text{B},\text{In}} - R_{i,n}^{\text{B},\text{Act}}) \eta_{i,n}^{\text{B},\text{Idle}} \big)}
				{p_{i,n}^{\text{B}} + q_{i,n}^{\text{B}}}, & 0 \leq R_{i,n}^{\text{B},\text{Act}} \leq R_{i,n}^{\text{B},\text{In}} \\[10pt]    	
				\frac{\mathsf{bid}'_{i,n} \times R_{i,n}^{\text{B},\text{In}} - R_{i,n}^{\text{B},\text{Est-}} p_{i,n}^{\text{B}} - \lambda \Delta t_n \big( R_{i,n}^{\text{B},\text{In}} \eta_{i,n}^{\text{B},\text{Use}} \big)}
				{p_{i,n}^{\text{B}} + q_{i,n}^{\text{B}}}, &  R_{i,n}^{\text{B},\text{In}} < R_{i,n}^{\text{B},\text{Act}} < R_{i,n}^{\text{B},\text{In}} + R_{i,n}^{\text{B},\text{Est-} }
			\end{cases}
		\end{align}
          \rule{\linewidth}{0.75pt}
		\begin{align}\label{scale3}
			{R}_{i,n}^{\text{B},\text{Est}} - R_{i,n}^{\text{B},\text{Act}} >
			\begin{cases}
				\frac{\mathsf{bid}'_{i,n} \times R_{i,n}^{\text{B},\text{Act}} - ({R}_{i,n}^{\text{B,Est}} - R_{i,n}^{\text{B},\text{In}}) p_{i,n}^{\text{B}} - \lambda \Delta t_n \big( R_{i,n}^{\text{B},\text{Act}} \eta_{i,n}^{\text{B},\text{Use}} - (R_{i,n}^{\text{B},\text{In}} - R_{i,n}^{\text{B},\text{Act}}) \eta_{i,n}^{\text{B},\text{Idle}} \big)}
				{p_{i,n}^{\text{B}} + q_{i,n}^{\text{B}}}, &  0 \leq R_{i,n}^{\text{B},\text{Act}} \leq R_{i,n}^{\text{B},\text{In}} \\[10pt] 	
				\frac{\mathsf{bid}'_{i,n} \times R_{i,n}^{\text{B},\text{In}} - ({R}_{i,n}^{\text{B,Est}} - R_{i,n}^{\text{B},\text{In}}) p_{i,n}^{\text{B}} - \lambda \Delta t_n \big( R_{i,n}^{\text{B},\text{In}} \eta_{i,n}^{\text{B},\text{Use}} \big)}
				{p_{i,n}^{\text{B}} + q_{i,n}^{\text{B}}}, &  R_{i,n}^{\text{B},\text{In}} < R_{i,n}^{\text{B},\text{Act}} < {R}_{i,n}^{\text{B},\text{Est}}
			\end{cases}	
		\end{align}
          \rule{\linewidth}{0.75pt}
		\begin{align}\label{scale4}
			R_{i,n}^{\text{B},\text{Act}}
			<
			\frac{{R}_{i,n}^{\text{B},\text{Est}} - 
				\begin{cases}
					\frac{- ({R}_{i,n}^{\text{B},\text{Est}} - R_{i,n}^{\text{B},\text{In}}) p_{i,n}^{\text{B}} + \lambda \Delta t_n (R_{i,n}^{\text{B},\text{In}} - R_{i,n}^{\text{B},\text{Act}}) \eta_{i,n}^{\text{B},\text{Idle}} }
					{p_{i,n}^{\text{B}} + q_{i,n}^{\text{B}}}, & 0 \leq R_{i,n}^{\text{B},\text{Act}} \leq R_{i,n}^{\text{B},\text{In}} \\[10pt] 	
					\frac{- ({R}_{i,n}^{\text{B},\text{Est}} - R_{i,n}^{\text{B},\text{In}}) p_{i,n}^{\text{B}} - \lambda \Delta t_n R_{i,n}^{\text{B},\text{In}} \eta_{i,n}^{\text{B},\text{Use}}}
					{p_{i,n}^{\text{B}} + q_{i,n}^{\text{B}}}, & R_{i,n}^{\text{B},\text{In}} < R_{i,n}^{\text{B},\text{Act}} \leq {R}_{i,n}^{\text{B},\text{Est}} 
				\end{cases}
			}
			{1 + \frac{\mathsf{bid}'_{i,n} - \lambda \Delta t_n \eta_{i,n}^{\text{B},\text{Use}} }{p_{i,n}^{\text{B}} + q_{i,n}^{\text{B}}}}
		\end{align}
          \rule{\linewidth}{0.75pt}
	\end{figure*}
	Let the right-hand side of \eqref{scale4} be  $\mathbb{R}$ for notational simplicity, we rewrite \eqref{probility} as 
	\begin{equation}
		\Pr(R_{i,n}^{\text{B},\text{Act}}<\mathbb{R})< \xi^\text{B}_{i,n}.
	\end{equation}
	Since the predicted values of $R_{i,n}^{\text{B},\text{Act}}$, derived from historical data, vary across different ESs and are not identically distributed, simplifying \eqref{probility} becomes analytically intractable. Fortunately, our subsequent simulations (e.g., Fig. \ref{fig:one_BU}) will demonstrate that, in the vast majority of cases, buyer utilities remain non-negative. Furthermore, within the context of our study, the proposed mechanism is primarily intended to facilitate the reutilization of surplus resources from ESs to generate additional revenue. Given that the penalties for default are relatively low, the theoretical probability of buyers experiencing negative utility is minimal.
	
	\noindent
	$\bullet$ \textit{For sellers}, executing LAContracts during practical transactions consistently results in positive utility. As demonstrated in Alg. 2, sellers always transact at prices equal to or higher than their asking prices when contracts are signed. In cases where a buyer breaches the contract, the associated standby energy consumption of the reserved resources is compensated by the buyer’s penalty payment. Additionally, the seller retains the option to reallocate the unused resources to serve its own users (e.g., vehicles), further enhancing its utility. Therefore, the overall profit for sellers remains strictly non-negative.
\end{proof}
\begin{lemma}\label{lemma:1.3}
	Implementing the pre-signed LAContracts during practical transactions (i.e., from $t_1$ to $t_N$) can hold the profit of ESs higher than without participating in this market.
\end{lemma}
\begin{proof}
	\textit{For a buyer}, we can say that  when the following inequality  holds,
	\begin{align}\label{bds}
		&\mathsf{bid}'_{i,n} \times \min(R^{\text{B},\text{Act}}_{i,n}, R^{\text{B},\text{In}}_{i,n}+
		R^{\text{B},\text{Tra}}_{i,n}) 
		\notag\\
		&- (R^{\text{B},\text{Tra}}_{i,n} - \kappa) p_{i,n}^\text{B} - \kappa q_{i,n}^\text{B}-\lambda \mathcal{E}^\text{B}_{i,n}  \notag\\ &<\mathsf{bid}'_{i,n} \times \min(R^{\text{B},\text{Act}}_{i,n}, R^{\text{B},\text{In}}_{i,n}),   
	\end{align}
	participating in the market may lead to degradation of profit. Similar to Lemma \ref{thm:theorem2}, this calls for proving that the risk shown by \eqref{risk2} can be controlled.
	\begin{equation}\label{risk2}
		\Pr(R_{i,n}^{\text{B},\text{Act}}<\mathbb{R}^\prime)< \xi^{\text{B}\prime}_{i,n}.
	\end{equation}
	In \eqref{risk2}, $ \xi^{\text{B}\prime}_{i,n}$ denotes  a positive tunable threshold. Similarly, we will demonstrate through simulations (e.g., Fig. \ref{fig:one_BU}, Fig. \ref{fig:half_BU} and Fig. \ref{fig:numerical_1} in Section \ref{5.1}) that buyers will always be encouraged to take part in the considered market thanks to higher utilities they can obtain, since the close-form of \eqref{risk2} is challenging to get due to the dynamic and uncertain nature of the market.
	
	\textit{As for sellers,} recall our earlier discussions: they prioritize transactions with buyers represented by other ESs, while placing their own covered users (i.e., vehicles) as secondary. Importantly, these internal users do not engage in contractual agreements and thus cannot incur penalties. However, since the resource price offered to external ES buyers is typically lower than the rate charged to internal users, inaccurate predictions of future demand may result in suboptimal allocations, potentially reducing the seller's overall profit. 
	
	To facilitate analysis, we present a simplified toy example. Consider a seller that owns 10 RBs during a specific time frame. Suppose its local (internal) resource demand—originating from its own vehicles—is predicted to be 5 RBs. Based on this prediction, the seller can allocate the remaining 5 RBs to external buyers (i.e., other ESs) via the auction mechanism. However, if serving internal demand generates higher revenue per RB compared to selling to external buyers, participating in the transaction may reduce the seller's overall profit.
	For instance, assume the actual internal demand turns out to be $8$ RBs. If the seller does not participate in the transaction, it could allocate all $8$ RBs to its own users, resulting in a total revenue of \(8 \times a\), where 
	$a$ is the unit price from internal users. In contrast, if it commits $5$ RBs to external buyers at a unit price $b$ (with $a \geq b$) and the remaining $5$ RBs to internal users, the total revenue becomes $5\times a + 5 \times b$. Clearly, if $8 \times a \geq 5 \times a + 5 \times b$, then the seller would have achieved higher revenue by not participating in the transaction.
	Therefore, according to above analysis, we need to have the following \eqref{eq39} holds.
	\begin{align}\label{eq39}
		&\mathbb{E} \left[ \mathcal{U}_{j,n}^{\text{S}} \right] \geq 
		\mathbb{E} \left[ U_{j,n}^{\text{S},1} \right] + 
		\mathbb{E} \left[ U_{j,n}^{\text{S},2} \right] + 
		\mathbb{E} \left[ U_{j,n}^{\text{S},3} \right] + 
		\mathbb{E} \left[ U_{j,n}^{\text{S},4} \right] \notag \\ 
		&\geq R_{j,n}^{\text{S},\text{Tra}} q_{j,n}^{\text{S}} + \sum_{\ell=0}^{R_{j,n}^{\text{S},\text{In}} - R_{j,n}^{\text{S},\text{Tra}}} 
		P_{j,\ell,n}^{\text{S}} \left( -\lambda  \mathcal{E}_{j,n}^{\text{S}} \Big| _{R_{j,n}^{\text{S},\text{Act}} = \ell} \right) \notag \\ 
		&\quad + v_j^{\text{S}} \min \left( R_{j,n}^{\text{S},\text{In}} - R_{j,n}^{\text{S},\text{Tra}}, R_{j,n}^{\text{S},\text{Act}} \right) \geq v_j^{\text{S}} \min \left( R_{j,n}^{\text{S},\text{In}}, R_{j,n}^{\text{S},\text{Act}} \right)
	\end{align}
	Due to market uncertainties, including price fluctuations, the uncertainty in each buyer's resource usage corresponding to a seller, and the dynamic nature of the seller's own resource utilization, ensuring that every seller achieves consistently high utility is extremely complex and impossible to prove rigorously. Instead, in experiments (e.g., Fig. \ref{fig:one_SU}, Fig. \ref{fig:half_SU} and Fig. \ref{fig:numerical_2} in Section \ref{5.1}) using both real-world datasets and numerical simulation, we can demonstrate that with controllable prediction errors in \eqref{p1}, sellers' profits can still be ensured.
\end{proof}
\begin{theorem}
	Our methodology is budget balance.
\end{theorem}
\begin{proof}Distinct from conventional auction mechanisms, this paper introduces a novel perspective by enabling a forward-looking approach that considers $N$-steps ahead in future transactions. This unique design necessitates careful examination of specific scenarios that may arise during practical implementations. Accordingly, to prove the theorem we analyze two representative cases, as formally presented in Lemmas \ref{lemma2.1} and \ref{lemma2.2}, the results of which conclude the proof.
\end{proof}
\begin{lemma}\label{lemma2.1}
	Our methodology is budget balance when there is no contract breaking events.
\end{lemma}
\begin{proof}
	Regard time frame $t_n$ as an example, according to Alg. 2, we have  $\mathsf{ask}_{j,n} \leq p_{i,j,n} \leq \mathsf{bid}_{i,j,n}$. The unit price for ESs $p^\text{S}_{j,n}$ and  $p^\text{B}_{i,n}$ are derived from $p_{i,j,n}$, and do not alter the financial inflow or outflow of the auctioneer. In the absence of any contract default, the total payments made by buyers are exactly equal to the total revenues received by sellers. As a result, the auctioneer operates in a budget-balanced manner and does not incur any financial loss within the market.
\end{proof}
\begin{lemma}\label{lemma2.2}
	Our methodology is budget balance when some buyers break their contracts.
\end{lemma}
\begin{proof}
	In Alg. 2, for each unit of resource defaulted by a buyer, the penalty is set as the maximum standby energy consumption cost among all the sellers associated with that buyer. Each seller then receives compensation equal to their own standby energy consumption cost. As a result, the total penalty paid by the defaulting buyer may exceed the total compensation distributed to the affected sellers. The surplus is retained by the auction platform as a service fee. Even under these circumstances, the auctioneer does not incur any loss, ensuring the market remains budget-balanced.
\end{proof}
Based on the above discussion, we can conclude that each individual transaction within a given time frame maintains budget balance. Consequently, our proposed method ensures budget balance is preserved throughout the entire transaction process.
\begin{theorem}
	Our methodology can support truthfulness for ESs.
\end{theorem}
Lemmas \ref{lem:3.1} and \ref{lem:3.2} provide two complementary perspectives for buyers and sellers, respectively, demonstrating that ESs are incentivized to report truthful information within our market.
\begin{lemma}\label{lem:3.1}
	Buyers in our considered market are truthful. 
\end{lemma}
\begin{proof}
	To facilitate the analysis of truthfulness regarding buyers, we consider the following two cases.
	
	\noindent
	$\bullet$ \textit{Case 1. }A buyer submits an untruthful bid that is larger than its true valuation, i.e., $\widehat{\mathsf{bid}}_{i,j,n} > \mathsf{bid}_{i,j,n}$. Let $\widehat{\mathcal{U}}^{\text{B}}_{i,n}$ denote the utility of buyer $\bm{b}_i$ in time frame $t_n$ under this untruthful bid, and $\widetilde{\mathcal{U}}^{\text{B}}_{i,n}$ be when bidding truthfully. Note that when the buyer places a bid in time frame $t_0$, their likelihood of eventual default remains highly uncertain. To reflect the buyer’s subjective assessment, we define  $\widetilde{\mathcal{U}}^{\text{B}}_{i,n}$ the maximum expected utility from the buyer’s own perspective. Under truthful bidding (see Alg. 2), the buyer can still be matched with the same seller, as the final transaction price is determined solely by the bids of losing buyers who satisfy the seller’s acceptance criteria. Consequently, the final payment made by the buyer remains unaffected by any deviation from truthful bidding. Therefore, we have $	\widehat{\mathcal{U}}^{\text{B}}_{i,n} = \widetilde{\mathcal{U}}^{\text{B}}_{i,n}$. 
	
	\noindent
	$\bullet$
	\textit{Case 2. }A buyers reports an untruthful bid that is lower than its true value, i.e., $\widehat{\mathsf{bid}}_{i,j,n} < \mathsf{bid}_{i,j,n}$. In this case, if the buyer is still matched with the same seller, their payment remains unchanged, implying $\widehat{\mathcal{U}}^{\text{B}}_{i,n} = \widetilde{\mathcal{U}}^{\text{B}}_{i,n}$. Conversely, if the buyer is no longer matched with the corresponding seller due to misreporting, their utility drops to zero, yielding $\widehat{\mathcal{U}}^{\text{B}}_{i,n} < \widetilde{\mathcal{U}}^{\text{B}}_{i,n}$. 
	
	Based on the above analysis, buyers have no incentive to misreport their bids, as doing so cannot lead to a higher utility. Therefore, our methodology induces truthful bidding behavior among buyers.
	%
	%
	%
	%
	%
\end{proof}
\begin{lemma}\label{lem:3.2}
	Sellers in our considered market are truthful.
\end{lemma}
\begin{proof}
	To support the analysis of truth-telling behavior among sellers, we examine the following two cases.
	
	\noindent
	$\bullet$
	\textit{Case 1. }Consider a seller who submits an untruthful ask that exceeds their true valuation, i.e., $\widehat{\mathsf{ask}}_{j,n}>\mathsf{ask}_{j,n}$. Let $\widehat{\mathcal{U}}^{\text{S}}_{j,n}$ denote the utility of seller $\bm{s}_j$ in time frame $t_n$ under the untruthful ask, and $\widetilde{\mathcal{U}}^{\text{S}}_{j,n}$ be the utility under truthful asking. Although the unit price per resource may increase slightly in certain cases, if the seller remains matched with a buyer, the number of buyers matched and the total allocated resources, denoted $\widehat{R}^{\text{S},\text{Tra}}_{j,n} $ will be less than or equal to $R^{\text{S},\text{Tra}}_{j,n} $, or even $\widehat{R}^{\text{S},\text{Tra}}_{j,n} = 0$. Thus, the seller's utility tends to decrease. According to Alg. 2, for the seller to remain matched, the inflated asking price must be higher than the minimum bid among buyers who satisfy the seller's true valuation, yet still lower than the asking prices of competing sellers. Only under such narrow and infrequent conditions could the seller potentially achieve a marginal increase in utility. However, if the seller is no longer matched with any buyer, their utility drops to zero. Therefore, in most cases, we have  $\widehat{\mathcal{U}}^{\text{S}}_{j,n} \leq \widetilde{\mathcal{U}}^{\text{S}}_{j,n}$.
	
	\noindent
	$\bullet$
	\textit{Case 2. }Now consider the case where a seller reports an untruthful ask that is lower than its true valuation, i.e., $\widehat{\mathsf{ask}}_{j,n}<\mathsf{ask}_{j,n}$. If the seller remains matched with a buyer, and there exist other sellers offering lower prices that still satisfy the seller’s true valuation, the pricing mechanism in Alg. 2 will lead to a reduced transaction price due to the inclusion of additional losing buyers who meet the seller’s criteria. As a result, the seller's final utility under untruthful reporting, $\widehat{\mathcal{U}}^{\text{S}}_{j,n}$ is less than or equal to that under truthful reporting $\widetilde{\mathcal{U}}^{\text{S}}_{j,n}$.
	
	Based on the above analysis, sellers in our market have no incentive to misreport their asking prices, as such deviations generally do not lead to higher utility. While there may be rare exceptions due to the structural differences between our auction design and conventional mechanisms, the overall incentive structure supports truthful reporting. Therefore, the proposed methodology ensures truthfulness for sellers (at least in most cases).
\end{proof}
All in all, under the proposed algorithmic mechanism, sellers generally do not benefit from misreporting their prices, as such behavior typically does not lead to increased utility. However, due to the inherent uncertainty of advance contract commitments and the dynamic nature of real-world markets, we cannot fully rule out the possibility that misreporting may occasionally yield favorable outcomes. Interestingly, this reflects a key distinction between our mechanism and conventional auction models. Unlike idealized settings, real markets inevitably involve some degree of strategic misreporting, and our design acknowledges and accommodates this complexity—underscoring its practical relevance and robustness.
\end{document}